\documentclass[aps,superscriptaddress, showpacs,preprintnumbers, superscriptaddress, nofootinbibt, twocolumn]{revtex4-1}
\usepackage{graphicx}
\usepackage{dcolumn}
\usepackage{bm}
\usepackage{caption}
\usepackage{subcaption}
\usepackage{epstopdf}
\usepackage{amsmath}
\usepackage{xcolor}
\usepackage{epsfig}
\usepackage{wasysym}
\usepackage{amssymb}
\usepackage{slashed}
\usepackage{color}
\usepackage{booktabs}
\usepackage{array}
\usepackage{tabularx}
\usepackage{multirow}
\usepackage{ragged2e}
\usepackage{epsfig}
\usepackage{braket,booktabs}
\usepackage{braket,amsmath}

\begin{document}


\title{Revisiting Chronology Protection Conjecture in The Dyonic Kerr-Sen Black Hole Spacetime}

\author{Teephatai Bunyaratavej}
\email{teephatai@gmail.com}
\affiliation{High Energy Physics Theory Group, Department of Physics,
Faculty of Science, Chulalongkorn University, Bangkok 10330, Thailand,}
\author{Piyabut Burikham}
\email{piyabut@gmail.com}
\affiliation{High Energy Physics Theory Group, Department of Physics,
Faculty of Science, Chulalongkorn University, Bangkok 10330, Thailand,}
\author{David Senjaya}
\email{davidsenjaya@protonmail.com}
\affiliation{High Energy Physics Theory Group, Department of Physics,
Faculty of Science, Chulalongkorn University, Bangkok 10330, Thailand,}

\date{\today}

\begin{abstract}
The Chronology Protection Conjecture (CPC) was first introduced by Hawking after his semi-classical investigation of the behaviour of a spacetime with closed timelike curves (CTCs) in response to scalar perturbations. It is argued that there would be instabilities leading to amplification of the perturbation and finally causing collapse of the region with CTCs. In this work, we investigate the CPC by exactly solving the Klein-Gordon equation in the region inside the inner horizon of the non-extremal Dyonic Kerr-Sen~(DKS) black hole, where closed timelike curves exist. Successfully find the exact radial solution, we apply the polynomial condition that turns into the rule of energy quantization. Among the quasi-resonance modes, only certain modes satisfy the boundary conditions of quasinormal modes~(QNMs). QNMs in the region inside the inner horizon of the rotating black hole with nonzero energy have only positive imaginary parts which describe states that grow in time. The exponentially growing modes will backreact and deform the spacetime region where CTC exists, hence the CPC is proven to be valid in the non-extremal Dyonic Kerr-Sen black hole spacetime. Since the Dyonic Kerr-Sen black hole is the most general axisymmetric black hole solution of the string inspired Einstein-Maxwell-dilaton-axion (EMDA) theory, the semiclassical proof in this work is also valid for all simpler rotating black holes of the EMDA theory. The structure of the Dyonic KS spacetime distinctive from the Kerr-Newman counterpart is also explored. 
\end{abstract}

\maketitle

\section{Introduction}
\label{sec:intro}
Investigation of various types of quantum fields in exterior black hole~(BH) spacetimes has been a booming research in the last decades. The explorations have uncovered numerous fascinating facets of quantum field theories in relation to the black hole physics. One of the examples is the quasi-resonance ringing that embodies unique black hole parameters that reverberates over spacetime when a black hole is perturbed. The quantized oscillation frequencies are termed quasinormal modes instead of normal modes, given their complex valued nature. The real component of a mode indicates the oscillation frequency, while the imaginary part signifies the damping or growing. These complex modes are important unique frequencies and directly relate to the  black hole parameters, i.e., mass, angular momentum and charges. 

After the groundbreaking detection of a binary black hole merger's gravitational wave signal on September 14, 2015, the Hawking radiation from an optical black hole analog was recently observed \cite{205,drori,nova}. This marks the emergence of black hole spectroscopy as a significant area of interest in physics. Quasibound states, quasinormal modes, and shadows of black holes are intriguing features of these cosmic entities, manifesting in observable spectra as particles enter the black hole. The importance of finding the exact solutions of the governing Klein-Gordon equation which describes the relativistic quantum mechanics of scalar fields in a given curved spacetime is that it will allow us to analytically investigate not only exact black hole's quasi-resonances, but also the black hole horizon thermodynamics that constitutes the first substantial step to understand the quantum field theory in curved spacetime. This is in contrast to the widely used WKB approximation, where the approximation breaks down for low valued momentum, $p=\hbar k\approx 0$, and the approximated wave function breaks down near potential barriers \cite{Ashok,Boris}.

Nonetheless, the study of how relativistic fields behave in the black hole interior spacetime has only started to gain attention relatively recently \cite{Chesler,Zilberman,McMaken,McMaken1}. One of primary justifications on the significance of the investigation of the inner region of Dyonic Kerr-Sen black hole is to verify the Chronology Protection Conjecture (CPC). Since the Dyonic Kerr-Sen black hole is the most general axisymmetric black hole solution that incorporates all other axisymmetric black hole solutions such as Kerr and Kerr-Sen and also their axionic and dilatonic versions \cite{Casadio,Hsu,Garcia,Semiz}, verifying the CPC in the Dyonic Kerr-Sen spacetime is automatically verifying CPC in the other simpler axisymmetric black hole solutions. Dyonic KS spacetime also has very rich spacetime structure~\cite{Sakti:2022izj,Sakti:2023fcu}, many aspects of which do not exist in General Relativity counterpart, the Kerr-Newman spacetime. It is interesting to explore the differences with respect to CTC and singularity behaviour.  

The term ``CPC" was encoded by Hawking in 1992 by employing a semi-classical investigation of the stress tensor in the wormhole throat~\cite{Hawking}. The conjecture argues that there exists a physical mechanism capable of averting the formation of closed timelike curves, that the region inside the inner horizon of rotating black holes with CTCs exhibit instability, i.e., particles or fields entering the aforementioned region will experience repeating blueshift proportional to the winding number of trajectory around the CTC region. In Hawking's investigation \cite{Hawking} of a radiation beam entering one end of a wormhole while considering vacuum fluctuations, it reveals that the beam would naturally realign itself before reaching the opposite end of the wormhole. This phenomenon implies that the accumulation of radiation becomes sufficiently significant to cause the collapse of the wormhole. The investigation in Ref.~\cite{Shinkai,Novikov} reveals that perturbation of relativistic boson fields to a wormhole throat will cause bifurcation of horizons and either explodes to form an inflationary universe or collapses to a black hole depending on the total input energy, respectively, negative or
positive. However, uncertainty still surrounded Hawking's conclusion as semi-classical quantum field theory in curved space is not reliable for high energy densities or brief time intervals close to the Planck scale with non-negligible backreaction to the spacetime, necessitating a comprehensive theory of quantum gravity for precise verdict~\cite{Visser:2002ua}. 

On the other hand, the CTC also exists behind the Cauchy horizon of the charged rotating BH in General Relativity, the Kerr-Newman spacetime. It was argued that these CTCs, if they even exist, are safe since it is hidden by the Cauchy horizon. In this work, we consider CPC of the KSBH by studying the behaviour of test scalar field in the region inside the inner horizon of a black hole in fully relativistic quantum description. Firstly, the Klein-Gordon equation in the Dyonic Kerr-Sen black hole background is constructed. We then apply the separation ansatz to separate the radial part. Following the recent developments of the Heun-family special functions and their applications as novel exact solutions of the Klein-Gordon's radial equation in various black hole backgrounds \cite{Vier22,senjaya1,senjaya2,senjaya3,senjaya4,senjaya5,senjaya6,senjaya7,senjaya8}. In this work, we present exact solutions of the scalar fields in the Dyonic Kerr-Sen black hole's interior region in terms of the Confluent Heun functions. From the polynomial condition, we obtain energy quantization condition. By applying appropriate boundary conditions for the QNMs, we find that in the region with the CTC, outgoing massive and massless scalar fields in the analytically continued spacetime are unstable, characterized by positive imaginary part of the quantized energy, leading to the exponential growth in time. For massless case, with exact explicit formula of the QNMs, we can show that all nonzero-energy modes are unstable and exponentially growing in time. Inevitably, this leads to the ever increasing backreaction to the spacetime region behind the inner horizon, and destroy the spacetime region where the presence of CTC is allowed.
 
This work is organized as the following. In Section~\ref{SecKSBH}, we review the classical properties of Dyonic Kerr-Sen black hole, i.e., the metric, horizons, singularities. In Section~\ref{SecCTC}, we discuss the closed timelike curve. The contour plots of the CTCs and the metric functions for various scenarios of the Dyonic KSBH's parameters are presented. Detail derivations to obtain the exact radial solutions of the governing Klein-Gordon equation and the quantized energy levels for both massive and massless scalar fields are presented in Section~\ref{SecEOM} and \ref{SecQSR}. Graphical visualizations of the complex quasi-resonance frequencies for all massive modes are also shown. Growing QNMs with specific boundary conditions are demonstrated to exist in Section~\ref{secCPC}. We argue that they validate the CPC. Section~\ref{SecCON} concludes our work.

\section{The Dyonic Kerr-Sen Black Hole}  \label{SecKSBH}

The Dyonic Kerr-Sen black hole is the most general axisymmetric black hole solution of the string inspired Einstein-Maxwell-dilaton-axion (EMDA) theory of gravity, where the generalization of the Einstein-Maxwell sector is performed by introducing coupling between the Maxwell electromagnetic field tensor $F_{\mu\nu}$ with the scalar dilaton field $\xi$ and introducing the pseudo-scalar axion field $\phi$ coupled to the dilaton field $\xi$. The theory is given by the following effective action defined in 3+1 dimension spacetime \cite{Sen,Baner},
\begin{multline}
S_{EMDA}=\frac{1}{16\pi}\int\left[R-2\partial_\mu\xi\partial^\mu\xi\right. \\ \left.-\frac{1}{3}H_{\rho\sigma\delta}H^{\rho\sigma\delta}+e^{-2\xi}F_{\alpha\beta}F^{\alpha\beta}\right]\sqrt{-g}d^4x,
\end{multline}
where $R$ is the Ricci scalar and $g$ is the metric tensor determinant. The Maxwell electromagnetic field tensor $F_{\mu\nu}$ is defined as partial derivatives of the $U(1)$ gauge field $A_\mu$,
\begin{equation}
F_{\mu\nu}=\partial_\mu A_\nu-\partial_\nu A_\mu.
\end{equation}

$H^{\rho\sigma\delta}$ is the Kalb-Ramond field tensor that is written in terms of the pseudo-scalar axion field $\phi$ according to,
\begin{equation}
H_{\alpha\beta\delta}=\frac{1}{2}e^{4\xi}\varepsilon_{\alpha\beta\delta\gamma}\partial^\gamma \phi,
\end{equation}
that is responsible for the coupling between the pseudo-scalar axion and the dilaton. As a result, axion becomes an inherent hair of the rotating black hole instead of becoming a part of the accretion disk.

\subsection{The Metric}
The Dyonic Kerr-Sen black hole in Boyer-Lindquist coordinates is described by the following metric~\cite{Wu,Jana},
    \begin{multline}
        ds^2 = -\bigg[1-\frac{r_s(r-d)-r_D^2} {\rho^2}\bigg]\ dt^2 +\frac{\rho^2}{\Delta}dr^2 +\rho^2d\theta^2 \nonumber \\-  2 \frac{r_s(r-d)-r_D^2}{\rho^2}a\sin^2\theta\ dtd\phi +\sin^2\theta\ d\phi^2 \times  \nonumber \\ \bigg[r(r-2d)-k^2+a^2+\frac{r_s(r-d)-r_D^2}  {\rho^2}a^2\sin^2\theta\bigg],  \label{metric}
    \end{multline}
    where,
    \begin{gather}
        \rho^2 = r(r-2d)-k^2+a^2\cos^2\theta,\\
        \Delta = r(r-2d) - r_s(r-d)-k^2+a^2+r_D^2,
    \end{gather}
    where the new variables in above equations are related to the physical properties by,
    \begin{gather}
        r_D^2 = Q^2 + P^2,\label{rD2eq}\\ 
        k = \frac{2PQ}{r_s},\label{keq}\\ 
        d = \frac{P^2-Q^2}{r_s},\label{deq}\\ 
        a = \frac{J}{M},
    \end{gather}
    where the {parameters} of the black hole are mass $M$, angular momentum per unit mass $a=\frac{J}{M}$, electric charge $Q$, magnetic charge $P$, dilaton charge $d$, and axion charge $k$.

    The condition $\Delta {\rm =0}$ makes the $g_{rr}$ divergent, indicating one way surfaces, i.e., the black hole horizons. The horizons' positions are then obtained by solving the quadratic equation,
\begin{gather}
 \Delta {\rm =0=}r\left(r-2d\right)-r_s\left(r-d\right)-k^2+a^2+r^2_D,\\
 r_{\pm }=\frac{r_s}{2}+d\pm \sqrt{{\left(\frac{r_s}{2}\right)}^2+d^2+k^2-\left(a^2+r^2_D\right)}, \label{rpm}
\end{gather}
where $r_+$ and $r_-$ are  the outer and inner horizons, respectively. Thus, it is possible to rewrite the $\Delta$ as,
\begin{equation}
    \Delta = \left(r-r_+\right)\left(r-r_-\right).
\end{equation}

Remarkably, there are double ellipsoidal singularities at $\rho^{2}=0$ or $r=r_{p,m}$ where
\begin{equation}
r_{p,m} = d\pm\sqrt{d^{2}+k^{2}-a^{2}\cos^{2}\theta},
\end{equation}
respectively. These are also the zeroes of $g_{rr}$.

One can also derive the metric determinant $g={\rm det}(g_{\mu\nu})$ and the metric inverse $g^{\mu\nu}$. After some straightforward algebras we find,
\begin{equation}
     g=-\rho^4{{\sin }^{{\rm 2}} \theta\ },   
\end{equation}
and,
\begin{gather}
g^{\mu\nu}=\left( \begin{array}{cccc}
-\frac{f^{\phi\phi}}{\Delta } & 0 & 0 & \frac{g_{0\phi}}{\Delta \sin^2 \theta} \\
0 & \frac{\Delta }{\rho^2} & 0 & 0 \\
0 & 0 & \frac{1}{\rho^2} & 0 \\
\frac{g_{0\phi}}{\Delta \sin^2 \theta} & 0 & 0 & {\frac{1}{\Delta {\sin }^{{\rm 2}}\theta} \left(1-\frac{r_s\left(r-d\right)-r^2_D}{\rho^2}\right)\ } \end{array}
\right), \label{metricinverse}\\
f^{\phi\phi}=r\left(r-2d\right)-k^2+a^2+\frac{r_s\left(r-d\right)-r^2_D}{\rho^2}a^2{{\sin }^{{\rm 2}} \theta\ }.
\end{gather}

\section{The Closed Timelike Curves}  \label{SecCTC}

Closed Timelike Curve (CTC) is a closed 1-dimensional curve in 4-dimensional spacetime where each tangent vector to the curve at every point is timelike. CTC has been found in a certain region of some well-known spacetime such as Tipler rotating cylinder~\cite{Tipler}, G\"odel universe~\cite{Godel}, Gott's time machine~\cite{Gott}, Wheeler wormholes~\cite{Wheeler}, Alcubierre's warp drive spacetimes~\cite{Alcub}, rotating cosmic string~\cite{Del}, axially symmetric spacetime with pure radiation~\cite{Sarma:2013zza} and locally AdS~\cite{Ahmed:2016fwj}, Kerr and Kerr-Newman black hole~(KNBH)~\cite{Chandrasekhar}, several analog gravity systems~\cite{Barcelo} and in various modified gravity scenarios~\cite{Franco,Santos,Silva,Nasci}. In the case of subextremal rotating black holes, the CTC is found inside the inner horizon of the Kerr spacetime while for the case of the electrically charged Kerr-Newman black hole, the CTCs exist both in the region inside the inner horizon and between its two horizons~\cite{Andre}. In this section, we will investigate in which region do CTCs exist in the Dyonic Kerr-Sen spacetime.

Let us consider the basis vector $\partial_\phi$. Since the $\phi$ is periodic, if the basis vector $\partial_\phi$ is timelike for any constant $t = t_0, r=r_0$ and $\theta=\theta_0$, then a circle (1-sphere) specified by a set of points $\{x\ |\ t = t_0, r=r_0,\theta=\theta_0\}$ describes a CTC. The region where CTCs exist can be found by investigating the metric component $g_{\phi\phi}$, i.e., whenever $g_{\phi\phi}<0$, the basis vector $\partial_\phi$ will be timelike which indicates the existence of CTCs.

The $g_{\phi\phi}$ component has the form of fourth degree polynomials in the numerator, but it also contains a point of constant radius at $\rho^{2}=0$ in the denominator where the function discontinues between positive infinity and negative infinity, the singularities. Further inspection shows that there are at most 6 CTC boundaries, characterized by four roots $r_{1},r_{2},r_{3},r_{4}$ of the numerator of $g_{\phi\phi}=0$, and two radii $r_p, r_m$ of the singularities. 

In the equatorial plane $\theta = \pi/2$, one of the roots~($r_{1}-r_{4}$) is simply $r_{p}$ and cancelled with the zero of $\rho^{2}=0$ in the denominator. This implies that there are at most 4 boundaries, three ($r_1, r_2, r_3$) are roots of $g_{\phi\phi}=0$ and one ($r_m$) is a singularity. Their expressions are given in Eqn.~(\ref{r1eq})-(\ref{r4eq}). 

\begin{widetext}
\begin{gather}
r_1 = \frac{1}{3r_s^3}\bigg[2\left(P^2-2 Q^2\right) r_s^2+\frac{X_1}{X_2^{1/3}}+X_2^{1/3}\bigg],\label{r1eq}\\
r_2 = \frac{1}{3r_s^3}\bigg[2\left(P^2-2 Q^2\right) r_s^2+\frac{\left(-1+i\sqrt{3}\right)}{2}\frac{X_1}{X_2^{1/3}}-\frac{\left(1+i\sqrt{3}\right)}{2}X_2^{1/3}\bigg],\\
r_3 = \frac{1}{3r_s^3}\bigg[2\left(P^2-2 Q^2\right) r_s^2-\frac{\left(1+i\sqrt{3}\right)}{2}\frac{X_1}{X_2^{1/3}}+\frac{\left(-1+i\sqrt{3}\right)}{2}X_2^{1/3}\bigg],\\
r_m = -\frac{2Q^2}{r_s},\label{r4eq}\\
X_1 = r_s^4 \left[4 \left(P^2+Q^2\right)^2-3 a^2 r_s^2\right],\\
X_2 = 8\left(P^2+Q^2\right)^3
r_s^6-9  \left(P^2+Q^2\right)a^2r_s^8-\frac{27 }{2}a^2 r_s^{10}+X_3,\\
X_3 = \frac{3\sqrt{3}}{2}r_s^8\sqrt{a^2\left[4 \left(a^2+9 \left(P^2+Q^2\right)\right)a^2  r_s^2-4 \left(P^2+Q^2\right)^2
\left(a^2+8 \left(P^2+Q^2\right)\right)+27 a^2 r_s^4\right]}
\end{gather}
\end{widetext}

In Fig.~\ref{FIG1}-\ref{FIG4}, the contour plots of the four boundaries $r_{1},r_{2},r_{3},r_{m}$ are shown. The plots exhibit the numerical values of the boundaries at each set of parameters $a,P,Q$ where the mass parameter is fixed at $r_s=2$. The white regions in the plots indicate the solution is complex and there will be no physical CTC boundary at those parameters' values. From the three roots $r_{1}-r_{3}$, there is always only one real root for small-charge cases and all three of them are real in large-charge case. On the other hand, $r_m$ is always real. This is why the boundaries of the CTC in Dyonic KSBH appear in pair in equatorial plane, hence create a bounded region(s) of CTC and there is no CTC at either positive or negative infinity of radial coordinate. It must be noted here that the index of the solutions does not indicate the radial order of them. For example, at low electric and magnetic charges, $r_1$ is the innermost boundary, where at higher charges it becomes the outermost boundary. 

In Fig.~\ref{FIG5}-\ref{FIG6}, the contour plots of the two horizons are shown in the same manner to the four boundaries. The white region again shows the set of parameters where the roots are complex and there is no horizon, i.e., the superextremal case. The two horizons have the same extremal limit that can be seen from the term in the square root of Eqn.~(\ref{rpm}).
    Compared to Kerr-Newman BH, the Dyonic Kerr-Sen BH has an interesting unique property which does not exist in Kerr-Newman BH. Consider initially a Dyonic Kerr-Sen BH with no electric and magnetic charge ($P,Q = 0$), but having small enough angular momentum to be a subextremal BH. Continuously increasing the charges will result in the BH crossing the extremal limit and becomes superextremal. This is presented by the white region in Fig.~\ref{FIG5},\ref{FIG6} where both horizons vanish. However, if we continue increasing the charges, the BH will again cross the extremal limit and return to be subextremal, in contrast to the KNBH where further increase of charges only results in superextremal spacetime or naked singularity.
    
    Another property of KSBH is that both the horizons can have negative value of $r$, this is shown in Fig.~\ref{FIG5},\ref{FIG6} in the `blue' region where the electric charge is large and the magnetic charge is small. This signifies the BH can have naked singularities even when it is subextremal, this is further discussed in the following section.

\subsection{Small and Large charge KSBH: Naked singularity solution}

\begin{figure*}[h!]
\centering
\includegraphics[width=15cm]{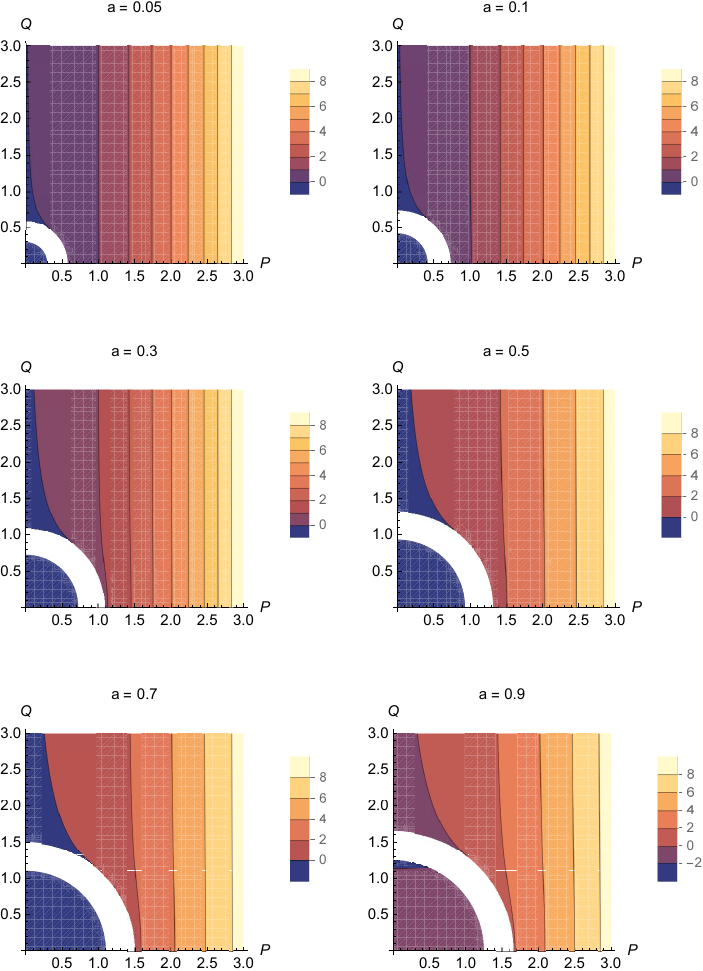}
\caption{Contour plot of the first boundary ($r_1$) of CTC regions.
} \label{FIG1}
\end{figure*}

\begin{figure*}[h!]
\centering
\includegraphics[width=15cm]{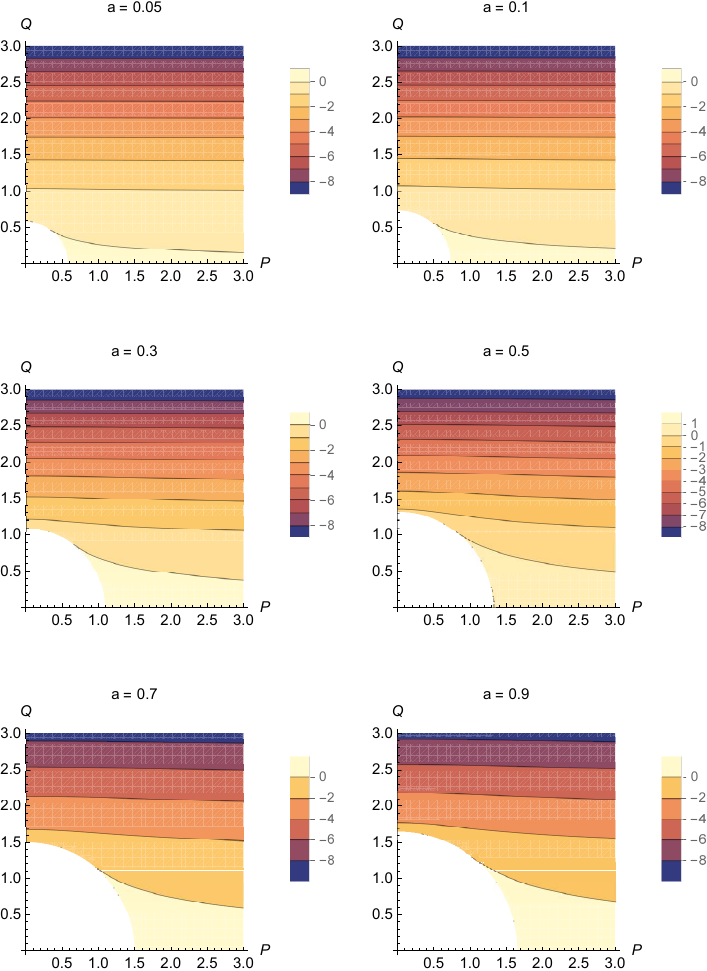}
\caption{Contour plot of the second boundary ($r_2$) of CTC regions.
} \label{FIG2}
\end{figure*}

\begin{figure*}[h!]
\centering
\includegraphics[width=15cm]{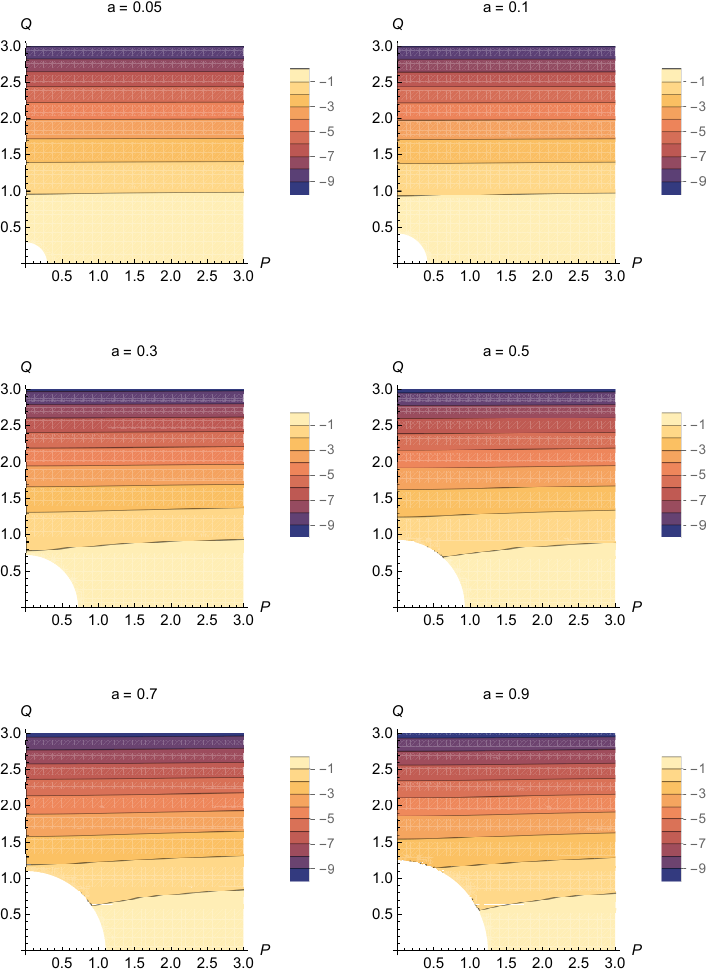}
\caption{Contour plot of the third boundary ($r_3$) of CTC regions.
} \label{FIG3}
\end{figure*}

\begin{figure*}[h!]
\centering
\includegraphics[width=15cm]{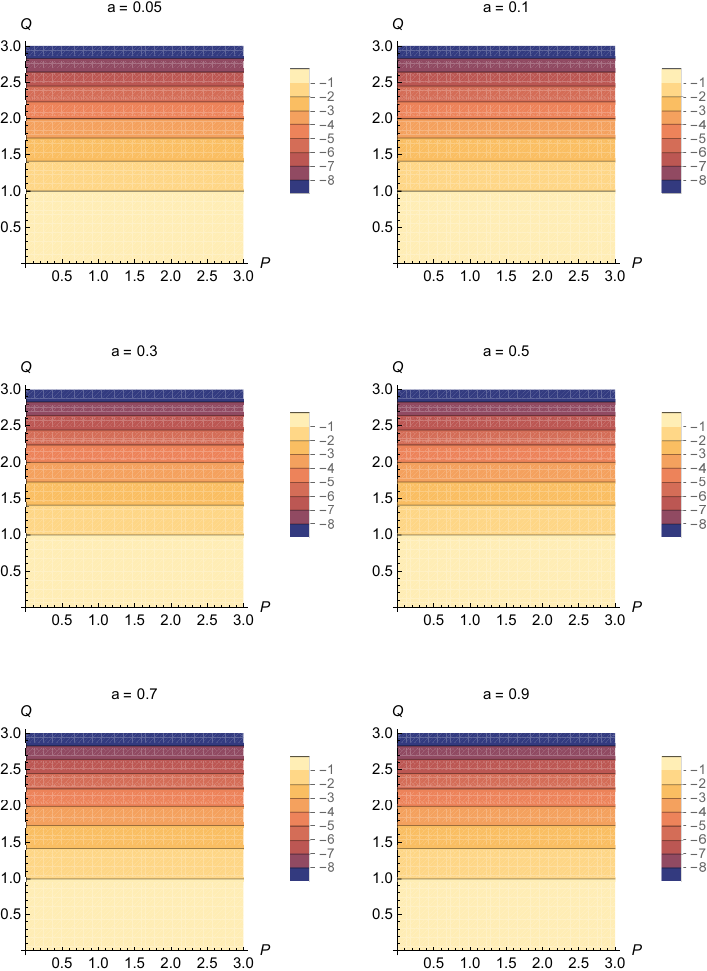}
\caption{Contour plot of the fourth boundary ($r_m$) of CTC regions.
} \label{FIG4}
\end{figure*}

\begin{figure*}[h!]
\centering
\includegraphics[width=15cm]{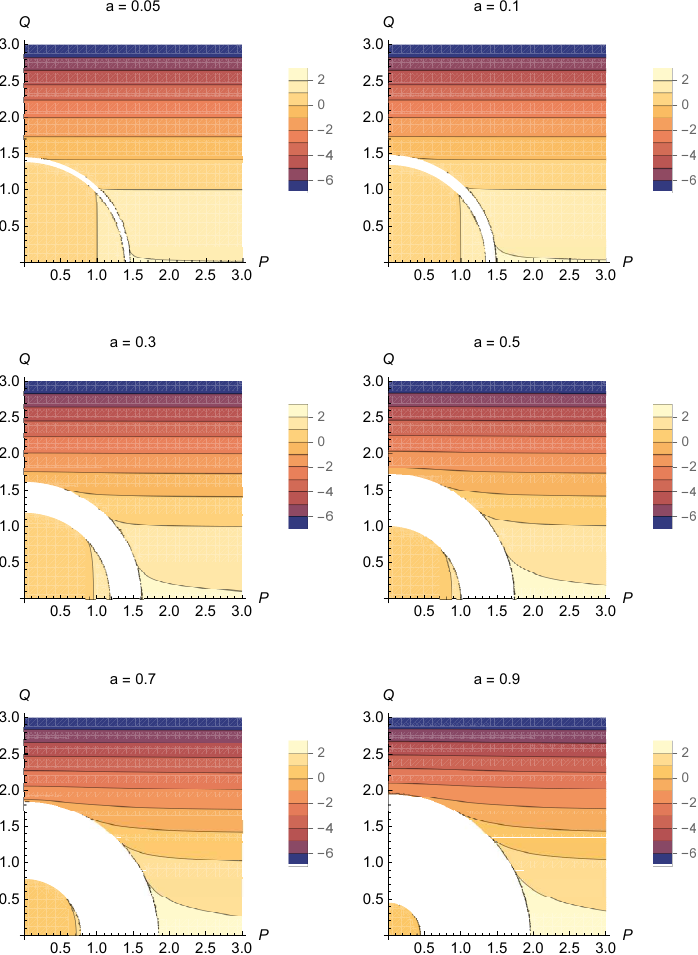}
\caption{Contour plot of the inner horizon ($r_-$).
} \label{FIG5}
\end{figure*}

\begin{figure*}[h!]
\centering
\includegraphics[width=15cm]{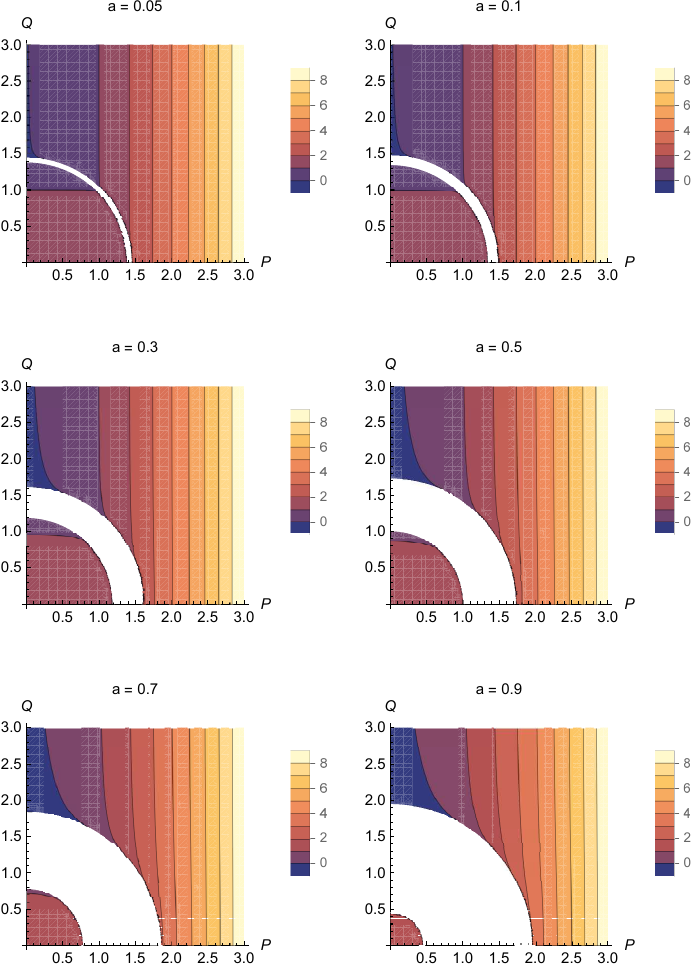}
\caption{Contour plot of the outer horizon ($r_+$).
} \label{FIG6}
\end{figure*}

\begin{figure*}[h!]
\centering
\includegraphics[width=15cm]{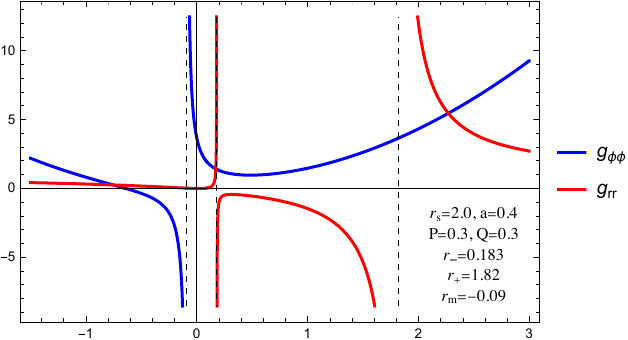}
\caption{Plot of $g_{rr}$ and $g_{\phi\phi}$ for very small-charge case
} \label{FIG7}
\end{figure*}

\begin{figure*}[h!]
\centering
\includegraphics[width=15cm]{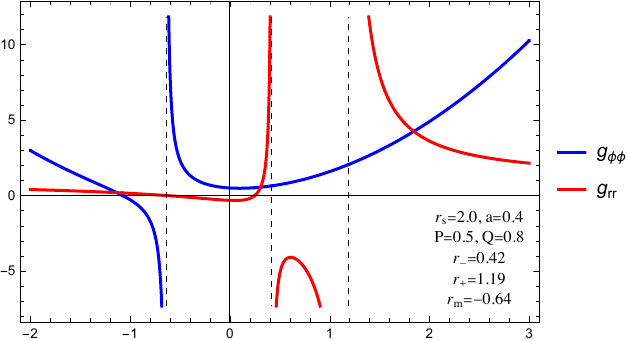}
\caption{Plot of $g_{rr}$ and $g_{\phi\phi}$ for small-charge case
} \label{FIG8}
\end{figure*}

\begin{figure*}[h!]
\centering
\includegraphics[width=15cm]{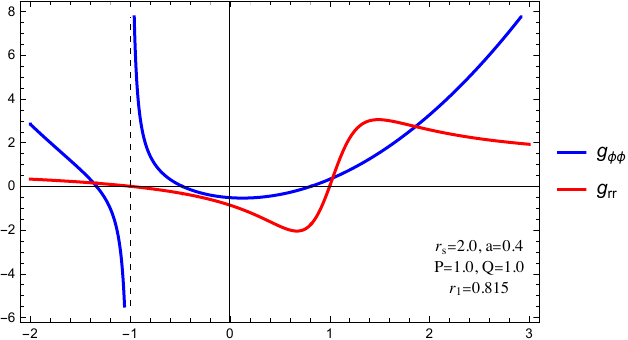}
\caption{Plot of $g_{rr}$ and $g_{\phi\phi}$ for moderate charge case
} \label{FIG9}
\end{figure*}

\begin{figure*}[h!]
\centering
\includegraphics[width=15cm]{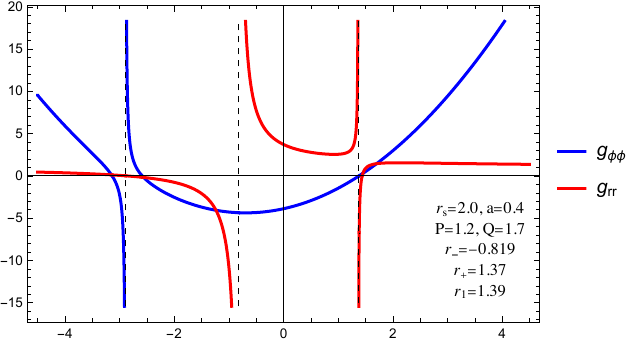}
\caption{Plot of $g_{rr}$ and $g_{\phi\phi}$ for large-charge case
} \label{FIG10}
\end{figure*}

    To obtain the full structure of spacetime including the singularities, CTC regions, and the two horizons, we plot the metric function $g_{rr}$ and $g_{\phi\phi}$ versus the radial distance $r$ for certain choices of parameters in Fig.~\ref{FIG7}-\ref{FIG10}. The singularities $r_{p,m}$ are at $g_{rr}=0$. The CTC boundary legended in each figure is the outermost boundary. It is found that in the very small-charge case, there are two CTC boundaries corresponding to $r_1$ and $r_m$ pair ~(see Fig.~\ref{FIG1}-\ref{FIG4}). For slightly larger charges (small-charge case), there are two CTC boundaries from $r_3$ and $r_m$. In both cases, $r_m$ is always the outer boundary as shown in Fig.~\ref{FIG7} and Fig.~\ref{FIG8}. The two cases are similar to the KNBH where all the CTCs are hidden behind the horizons, but for the moderate-charge and large-charge cases, this statement is no longer valid. For these cases~(Fig.~\ref{FIG10}), all 4 boundaries exist. 
    
    Both moderate-charge and large-charge spacetimes have naked singularity in the large $r$ region. In the large-charge case, the outermost CTC boundary goes beyond the outer horizon $r_+$, demonstrating that there exists a small CTC `island' outside the horizons. However, this occurs in the region of spacetime where $g_{rr}$ is negative, indicating that $\partial_r$ is also timelike and hence this `island' spacetime is unphysical. This behaviour only occurs in the large-charge case since the outermost surface of constant radius where $g_{rr}$ changes sign from positive to negative lies outside the outer horizon, as opposed to the smaller charge case where they coincides. For the moderate charge case where there is no horizon, generically there could be either 2 or 4 CTC boundaries as shown in the contour plots, Fig.~\ref{FIG1}-\ref{FIG4}.

    The large-charge KSBH such as depicted in Fig.~\ref{FIG10} behaves like an exotic antigravity spacetime in the large $r>r_{p}$ region with the presence of naked singularity at $r_{p}$. The metric function indicates gravity becomes {\it repulsive} in this region and it is infinitely repulsive at the singularity, i.e., nothing can reach the naked singularity from this side of spacetime. Interestingly, behind the horizon at $r_{+}<r_{p}$, there is a physical spacetime region $r_{-}<r<r_{+}$ which allows the existence of CTCs. However, observer in this region will see a cosmic horizon at $r_{+}$ and will never be in (classical) contact with the region $r>r_{p}$.  

One may do further analytical investigation on the conditions of Dyonic Kerr-Sen black hole's parameters and their relations to $g_{rr},g_{\phi \phi}$ to check whether CTC could be found in the physical spacetime outside the event horizon, $r_{+}$. First, by using (\ref{rD2eq})-(\ref{deq}), we can write $r_\pm$ and $\delta_r=r_{+}-r_{-}$ explicitly in terms of black hole parameters, $r_{s},a,Q,P$, 
\begin{gather}
r_\pm=\frac{r_s}{2}+\frac{P^2-Q^2}{r_s}\pm\frac{1}{2}\delta_r,\\
\delta_r=\sqrt{\frac{\left(r_s^2-2\left(P^2+Q^2\right)\right)^2}{r_s^2}-4a^2}. \label{dreq}
\end{gather}

Since we are interested in the case of subextremal black hole, $\delta_r$ must be larger than zero. This condition leads to two possibilities,
\begin{align}
    &1. \ \  P^2+Q^2-\frac{r_s^2}{2}<-r_s\sqrt{a^2}, \label{lowcharge}\\
    &2. \ \  P^2+Q^2-\frac{r_s^2}{2}>r_s\sqrt{a^2}, \label{highcharge}
\end{align}
where the case (1.) leads to a lightly charged black hole, where the electric charge must follow the inequality, $0<Q^2<\frac{r_s^2}{2}-P^2-r_s\sqrt{a^2}$, and the case (2.) leads to a highly charge black hole, where the electric charge must follow the inequality, $Q^2>\frac{r_s^2}{2}-P^2+r_s\sqrt{a^2}$. The case (2.) is the special feature of the Dyonic Kerr-Sen black hole where high ($Q$)-charge black hole solution is possible without a naked singularity. This is the `blue' region mentioned earlier in Fig.~\ref{FIG1}-\ref{FIG4}.

Let us consider the singularities at $\theta = \pi/2$, 
\begin{gather}
    r_p=\frac{2P^2}{r_s},\label{rp} \\ 
    r_m=-\frac{2Q^2}{r_s}, \label{rm}
\end{gather}
where it is clear that for general non zero $P^2>0$ and $Q^2>0$, $r_p$ will be positive and $r_m$ will be negative.

One can show that the condition $r_p>r_+$ leads to $a^{2}>0$,
\begin{gather}
  \frac{2P^2}{r_s}>\frac{r_s}{2}+\frac{P^2-Q^2}{r_s}+\frac{\delta_r}{2}, \label{one}
\end{gather}
substituting $\delta_r$ by (\ref{dreq}), clearly the condition is $a^2>0$. However, either the condition \eqref{lowcharge} or \eqref{highcharge} also needs to be satisfied. One can rearrange \eqref{one} so that,
\begin{equation}
P^2+Q^2-\frac{r_s^2}{2}>\frac{1}{2}r_s\delta_r>0,  
\end{equation}
which implies that $r_p>r_+$ happens only when \eqref{highcharge} is valid. So the black hole must be rotating and highly charged. Similarly, one can also check the condition for $r_m<r_-$ and find exactly the same condition as for $r_p>r_+$. Therefore, for large-charge Dyonic Kerr-Sen black hole, the following ordering holds,
\begin{equation}
    r_m<r_-<r_+<r_p.
\end{equation}

For small-charge Dyonic Kerr-Sen black hole, one may follow the same procedure and find that $r_p<r_{-}$ leads to the condition $a^2>0$ and \eqref{lowcharge}. Therefore the following ordering holds,
\begin{equation}
   r_m<r_p<r_-<r_+,
\end{equation}
for the small-charge KSBH. There is no possible black hole solution with $r_p$ between $r_-$ and $r_+$.

Now, let us consider $g_{\phi\phi}$ at $\theta = \pi/2$. After some rearrangement, one can express $g_{\phi\phi}$ as
\begin{gather}
g_{\phi\phi}=\frac{1}{\rho^{2}}\left[(\rho^{2}+a^2)^2-a^2\Delta \right], \label{gff}\\
\rho^{2}=\left(r-2\frac{P^2}{r_s}\right)\left(r+2\frac{Q^2}{r_s}\right),
\end{gather}
or, in the terms of $r_m,r_p$ at $\theta = \pi/2$, we obtain,
\begin{gather}
\rho^{2}=\left(r-r_p\right)\left(r-r_m\right),
\end{gather}

At the black hole horizons, $\Delta=0$ and $g_{\phi\phi}$ is simplified to
\begin{equation}
    g_{\phi\phi}(r_\pm)=\frac{\left(\rho^{2}+a^2\right)^2}{\left(r_\pm-r_p\right)\left(r_\pm-r_m\right)},
\end{equation}
after some lines of straightforward algebras, one finds,
\begin{gather}
g_{\phi\phi}(r_+)= -r_s^2  \frac{P^2+Q^2-\frac{r_s^2}{2}-\frac{r_s}{2}\delta_r}{P^2+Q^2+\frac{r_s^2}{2}+\frac{r_s}{2}\delta_r},\\
g_{\phi\phi}(r_-)= -r_s^2  \frac{P^2+Q^2-\frac{r_s^2}{2}+\frac{r_s}{2}\delta_r}{P^2+Q^2+\frac{r_s^2}{2}-\frac{r_s}{2}\delta_r}. \label{g33con2}
\end{gather}
The condition $g_{\phi\phi}(r_+)<0$ implies that 
\begin{equation}
P^{2}+Q^{2}>\frac{r_{s}^{2}}{2}+\frac{r_{s}}{2}\delta_{r},  \label{g33con}
\end{equation}
which leads to the simple condition $a^{2}>0$. However, using \eqref{lowcharge} and \eqref{highcharge}, we can conclude that \eqref{g33con} can only be true for \eqref{highcharge}, i.e., the large-charge BH. CTC can exist in the region $r\geq r_+$ if the angular momentum of the BH, $a$, has value in such a way that,
\begin{equation}
    0<a^2<\frac{1}{4r_s^2}\left[2\left(Q^2+P^2\right)-r_s^2\right]^2.
\end{equation}

Now, let us investigate the behaviour of $g_{\phi\phi}$ in the region $r>r_+$. The function $g_{\phi\phi}$ can be written in its most simplified form as
\begin{equation}
  g_{\phi\phi}=\frac{1}{r_s^2\left(rr_s+2Q^2\right)}F_\phi, \label{simp}
\end{equation}
where,
\begin{multline}
F_\phi=r_s\left[4Q^4r+2Q^2r_s(2r^2+a^2)\right.\\ \left.+r^2\left\{r^3+a^2(a+r_s)\right\}\right] -2P^2\left(rr_s+2Q^2\right)^2.
\end{multline}

Since the $ g_{\phi\phi}$ in \eqref{simp} can be thought as a multiplication of two functions, where the first term is clearly always positive in the region $r>r_m=-\frac{2Q^2}{r_s}$, we need further searching for extrema of $F_\phi$ by finding the root of $\partial_r F_\phi=0$. The resulting equation is a quartic equation that only has two distinct roots,
\begin{equation}
 r_{\phi\pm}=\frac{1}{3}\left[\frac{2P^2-4Q^2}{r_s}\pm\sqrt{\frac{4(P^2+Q^2)^2}{r_s^2}-3a^2}\right].   
\end{equation}

Next, we can estimate the general behaviour of $F_\phi$ from $r\to-\infty$ to $r\to\infty$. From $r\to-\infty$, $F_{\phi}$ increases until it reaches the local maxima at $r_{\phi-}$, then decreases until it reaches the local minima at $r_{\phi+}$. Then $F_{\phi}$ monotonically increases towards $+\infty$. One can verify that the local minima is always located inside the black hole outer horizon, $r_+>r_{\phi+}$ as the following,
\begin{multline}
r_+- r_{\phi+}=\frac{1}{6}\left[\frac{2(P^2+Q^2)}{r_s}+3(r_s+\delta_r)\right.\\ \left.-2\sqrt{\frac{4(P^2+Q^2)^2}{r_s^2}-3a^2}\right]>0.
\end{multline}

Therefore, for small-charge BH \eqref{lowcharge},  $g_{\phi\phi}(r_+)>0$ and since $g_{\phi\phi}$ is monotonically increasing in the region $r>r_+$, it is impossible to find CTC in the region $r>r_+$.

Now we consider the region around the inner horizon, $r=r_{-}$. For $r_{s}>\delta_{r}$~(small charge) KSBH, since $a^{2}>0$ implies that
\begin{equation}
P^{2}+Q^{2}<\frac{r_{s}^{2}}{2}-\frac{r_{s}}{2}\delta_{r},
\end{equation}
$g_{\phi\phi}(r_{-})$ is always positive and there is no CTC.

On the other hand for $r_{s}<\delta_{r}$~(large charge) KSBH, since $a^{2}>0$ implies that the numerator of \eqref{g33con2} is always positive and the denominator is always negative, $g_{\phi\phi}(r_{-})$ is always negative and there are CTCs in $r=r_{-}$ region. 

Examining $\partial_r g_{\phi\phi}$ reveals that there is an asymptote at $r_{m}=-\frac{2Q^2}{r_s}$, the inner singularity. The behaviour of $g_{\phi\phi}$ around the asymptote can be investigated as follows. First, let us make a Taylor expansion around $r=r_m$,
\begin{equation}
g_{\phi\phi} \approx a^2\left[\frac{r_s}{r-r_m}+1\right].
\end{equation}
As $r\to r_{m}$ from the right, $g_{\phi\phi}\to +\infty$, and as $r\to r_{m}$  from the left, $g_{\phi\phi}\to -\infty$. This indicates that there is always a region where CTC exists on the left side of $r=r_m$, which is always inside $r_-$.

For highly charged black hole, since $r_p>r_+$, we need to check the sign of $g_{\phi\phi}$ at $r_p$ to see whether CTC can exist in the physical exterior space. We found that 
\begin{equation}
    g_{\phi\phi}(r_p)=\frac{2a^2}{a\left(P^2+Q^2\right)}\left[P^2+Q^2+\frac{r_s^2}{2}\right]>0.
\end{equation}

Since $g_{\phi\phi}$ is a monotonically increasing function in the region $r\geq r_p$, therefore there is no CTC in the physical exterior space.

\section{The Klein-Gordon Equation}  \label{SecEOM}
    The Klein-Gordon equation is a covariant relativistic wave equation that describes the dynamics of a massive or massless spin-0 particle in a particular spacetime. The equation reads
    \begin{equation}
        \left[\nabla_\mu\nabla^\mu - \left(\frac{mc}{\hbar}\right)^2\right]\psi = 0,
    \end{equation}
    where $\nabla_\mu$ is the covariant derivative, $m$ is the rest mass of the scalar field $\psi$. As the covariant derivatives act to a scalar field $\psi$, they can be rewritten explicitly as the Laplace-Beltrami operator as
    \begin{equation}
        \left[\frac{1}{\sqrt{-g}}\partial_\mu\left(\sqrt{-g}\ g^{\mu\nu}\partial_\nu\right) - \left(\frac{mc}{\hbar}\right)^2\right]\psi = 0.
    \end{equation}
Note that the differential operator acts first to the field $\psi$ in the right-to-left order, $\partial(A\partial)\psi=\partial(A\partial \psi)$. 

Using the metric inverse \eqref{metricinverse}, we thoroughly construct the explicit form of each component of the Laplace-Beltrami operator in the Dyonic Kerr-Sen black hole,
\begin{widetext}
\begin{gather}
\frac{1}{\sqrt{-g}}{\partial }_0(\sqrt{-g}g^{00}{\partial }_0){\rm =-}\frac{{\rm 1}}{\Delta \rho^2}\left\{{\left[r\left(r-2d\right)-k^2+a^2\right]}^2-\Delta a^2{{\sin }^{{\rm 2}} \theta\ }\right\}{\partial }^2_{ct},\\
\frac{1}{\sqrt{-g}}{\partial }_0(\sqrt{-g}g^{03}{\partial }_3){\rm =-}\frac{\left[r\left(r-2d\right)-k^2+a^2-\Delta \right]a}{\Delta \rho^2}{\partial }_{ct}{\partial }_\phi,\\
\frac{1}{\sqrt{-g}}{\partial }_{{\rm 3}}(\sqrt{-g}g^{30}{\partial }_0){\rm =-}\frac{\left[r\left(r-2d\right)-k^2+a^2-\Delta \right]a}{\Delta \rho^2}{\partial }_{ct}{\partial }_\phi,\\
\frac{1}{\sqrt{-g}}{\partial }_{{\rm 1}}(\sqrt{-g}g^{11}{\partial }_1){\rm =}\frac{1}{\rho^2}{\partial }_r\left(\Delta {\partial }_r\right),\\
\frac{1}{\sqrt{-g}}{\partial }_{{\rm 2}}(\sqrt{-g}g^{22}{\partial }_2){\rm =}\frac{1}{\rho^2{\sin  \theta\ }}{\partial }_\theta\left({\sin  \theta\ }{\partial }_\theta\right),\\
\frac{1}{\sqrt{-g}}{\partial }_{{\rm 3}}(\sqrt{-g}g^{33}{\partial }_3){\rm =}\frac{\Delta -a^2{{\sin }^{{\rm 2}} \theta\ }}{\Delta {{\sin }^2 \theta\ }\rho^2}{\partial }^2_\theta,\\
\end{gather}
\end{widetext}
and after collecting all of the terms, we obtain the following differential equation,
\begin{multline}
\left[{\rm -}\frac{{\rm 1}}{\Delta \rho^2}\left\{{\left[r\left(r-2d\right)-k^2+a^2\right]}^2-\Delta a^2{{\sin }^{{\rm 2}} \theta\ }\right\}{\partial }^2_{ct}\right. \\ \left.-2\frac{\left[r\left(r-2d\right)-k^2+a^2-\Delta \right]a}{\Delta \rho^2}{\partial }_{ct}{\partial }_\phi\right. \\ \left.+\frac{1}{\rho^2}{\partial }_r\left(\Delta {\partial }_r\right)+\frac{1}{\rho^2{\sin  \theta\ }}{\partial }_\theta\left({\sin  \theta\ }{\partial }_\theta\right)\right. \\ \left.+\frac{\Delta -a^2{{\sin }^{{\rm 2}} \theta\ }}{\Delta {{\sin }^2 \theta\ }\rho^2}{\partial }^2_\phi\right]\psi-\frac{m^2 c^2}{{\hbar }^2}\psi=0. \label{fullwave}
\end{multline}

\subsection{Separation of Variables}
Due to the temporal and azimuthal symmetry, the separation ansatz is applied \cite{35},
\begin{gather}
\psi\left(t,r,\theta,\phi\right)=e^{{-}i\frac{E}{c}ct{+}im_\ell \phi}R\left(r\right)T\left(\theta\right).
\end{gather}

For the sake of notation simplicity, we define the following dimensionless variables, $\Omega=\frac{Er_s}{\hbar c}$ and $\Omega_0=\frac{E_0r_s}{\hbar c}$, where $E_{0}=mc^{2}$. Now, multiplying the whole equation by $\rho^2r^2/\psi \left(t,r,\theta ,\phi \right)$, we arrive at the equation,
\begin{multline}
\left[\frac{1}{T{\sin  \theta\ }}{\partial }_\theta\left({\sin  \theta\ }{\partial }_\theta T\right)-\frac{m^2_l}{{{\sin }^2 \theta\ }}\right. \\ \left.-\left(\frac{\Omega^2_0 a^2}{r^2_s}-\frac{\Omega^2a^2}{r^2_s}\right){{\cos }^2 \theta\ }\right]+\left[\frac{1}{R}{\partial }_r\left(\Delta {\partial }_rR\right)\right. \\ \left.+\frac{\Omega^2}{r^2_s}\frac{{\left(r\left(r-2d\right)-k^2+a^2\right)}^2}{\Delta }-\frac{\Omega^2a^2}{r^2_s}\right. \\ \left.-2\frac{\left(r\left(r-2d\right)-k^2+a^2-\Delta \right)a}{\Delta }\left(\frac{\Omega m_\ell }{r_s}\right)\right. \\ \left.+\frac{m^2_l a^2}{\Delta }-\frac{\Omega^2_0}{r^2_s}\left(r\left(r-2d\right)-k^2\right)\right]=0.
\end{multline}

Notice that since the terms inside the first square bracket depends only on $\theta$ while the rest are functions of $r$, they can be separated,
\begin{multline}
  \frac{1}{{T\sin  \theta\ }}{\partial }_\theta\left({\sin  \theta\ }{\partial }_\theta T\right)-\frac{m^2_l}{{{\sin }^2 \theta\ }}\\-\left(\frac{\Omega^2_0a^2}{r^2_s}-\frac{\Omega^2a^2}{r^2_s}\right){{\cos }^2 \theta\ }=-\lambda^{m_\ell }_\ell ,   
  \end{multline}
where $\lambda^{m_\ell }_\ell $ is a constant. For static spherically symmetric spacetime of non-rotating black holes, the separation constant $\lambda^{m_\ell }_\ell $ is exactly equal to $\ell\left(\ell+1\right)$. Therefore, the polar differential equation has exact solution in terms of the associated Legendre polynomial, $P^{m_\ell }_l (\cos \theta)$. 

With the presence of angular momentum, $a\neq 0$, the polar equation is generalized and has series solution named Spheroidal Harmonics, $S^{m_\ell }_l$. The Spheroidal Harmonics can be written as a series expansion of associated Legendre polynomial given by
\begin{gather}
T(\theta){\rm =}S^{m_\ell }_l\left(\sigma,{\cos  \theta\ }\right)=\sum^{\infty }_{r=-\infty }{d^{lm_\ell }_r\left(\sigma\right)P^{m_\ell }_{l+r}\left({\cos  \theta\ }\right)},
\end{gather}
where,
\begin{equation}
 \sigma= \frac{\Omega^2_0a^2}{r^2_s}-\frac{\Omega^2a^2}{r^2_s}.
\end{equation}

The coefficient $d^{lm_\ell }_r$ is amplitude of the mode $\left(l+r\right), m_\ell $. It is possible, for the case $|\sigma|<1$ to algebraically calculate $\lambda^{m_\ell }_\ell $ via perturbation theory  \cite{Press,Berti1,Berti2,Cho:2009wf,Suzuki:1998vy}, resulting in 
\begin{equation}
\lambda^{m_\ell }_\ell =l(l+1)-2\sigma\left(\frac{m_\ell ^2+l(l+1)-1}{(2l-1)(2l+3)}\right)+O\left(\sigma^2  \right).
\end{equation}

\subsection{The Radial Equation}
Now we consider the radial equation,
\begin{multline}
\partial_r^2 R+\left(\frac{1}{r-{r_-}}+\frac{1}{r-{r_+}}\right)\partial_r R\\+\frac{1}{\delta_r^2}\left[{\left(\frac{1}{r-{r_+}}-\frac{1}{r-r_-}\right)}^2\times\right. \\ \left.{\left\{\frac{\left(a^2-k^2+r\left(r-2d\right)-a m_\ell \right){\Omega}}{r_s}\right\}}^2\right. \\ \left.-\delta_r\left(\frac{1}{r-{r_+}}-\frac{1}{r-{r_-}}\right)\times\right. \\ \left.\left\{K^{m_\ell }_l+\frac{\left(a^2-k^2+r\left(r-2d\right)\right)\Omega_0^2}{r_s^2 }\right\}\right]R=0,
\end{multline}
where the constant $K^{m_\ell }_l$ is used to abbreviate the expression,
\begin{gather}
K^{m_\ell }_l=\frac{\Omega^2}{r^2_s}a^2-\frac{\Omega^2_0}{r^2_s}a^2-2\frac{\Omega m_\ell }{r_s}a+\lambda^{m_\ell }_\ell .
\end{gather}
   
Transforming the radial coordiante $r\to x=\frac{(r-\mathrm{r_-})}{{\delta}_{r}}$, we obtain the radial equation in terms of $x$,
\begin{multline}
\partial_{x}^2R+\frac{\left(2x-1\right)}{\left(x-1\right)x}\partial_{x} R+({T_1}^2+T_2)R=0,
\end{multline}
where we have defined,
\begin{multline}
T_1=\frac{1}{\delta_rr_s\left(x-1\right)x}\left[amr_s\right. \\ \left.-\left\{a^2-k^2+\left(r_-+\delta_rx\right)\left({r_-}-2d+\delta_rx\right)\right\}{\Omega}\right],    
\end{multline}
and,
\begin{multline}
T_2=-\frac{1}{r_s^2 {\left(x-1\right)}x}\left[K^m_\ell r_s^2 \right. \\ \left.+\left\{a^2-k^2+\left(r_-+\delta_rx\right)\left({r_-}-2d+\delta_rx\right)\right\}\Omega_0^2\right].
\end{multline}

By applying fractional decomposition to $T_1$, we obtain,
\begin{multline}
T_1=-\frac{K_1}{\delta_r x}+\frac{\delta_r{\Omega}}{r_s}\\-\frac{-K_1r_s+(r_-^2-2dr_--r_+^2+2dr_+){\Omega}}{\delta_r r_s\left(x-1\right)},    
\end{multline}
where we have defined,
\begin{equation}
K_1=-\frac{amr_s+\left(k^2-a^2+2dr_--r_-^2\right){\Omega}}{r_s}.
\end{equation}

Let us define $K_3$ as the coefficient of ${\left(x-1\right)}^{-1}$ as follows,
\begin{equation}
    K_3=-\frac{-K_1r_s+(r_-^2-2dr_--r_+^2+2dr_+){\Omega}}{\delta_r r_s},
\end{equation}
and after some lines of algebras, we can write $T_1^2$ as 
\begin{multline}
T_1^2=\frac{{K_3}^2}{{(x-1)}^2}+\frac{{K_1}^2}{\delta_r^2x^2}+\frac{\delta_r^2{{\Omega}}^2}{r_s^2 }\\+\frac{2K_1\left(K_3r_s-\delta_r{\Omega}\right)}{\delta_rr_sx}-\frac{2K_3\left(K_1r_s-\delta_r^2{\Omega}\right)}{\delta_rr_s\left(x-1\right)}.
\end{multline}

Now let us consider $T_2$, applying the fractional decomposition, we obtain,
\begin{multline}
T_2=\frac{K_2}{x}-\frac{\delta_r^2\Omega_0^2}{r_s^2 }\\+\frac{-K_2r_s^2+(r_-^2-2dr_--r_+^2+2dr_+)\Omega_0^2}{r_s^2\left(x-1\right)},
\end{multline}
where we have defined,
\begin{equation}
K_2=-\frac{-K_l^{m_\ell }r_s^2 +(k^2-a^2+2dr_--r_-^2)\Omega_0^2}{r_s^2 },
\end{equation}
and for the sake of simplicity, let us define $K_4$ as the coefficient of ${\left(x-1\right)}^{-1}$ as
\begin{equation}
K_4=\frac{-K_2r_s^2+(r_-^2-2dr_--r_+^2+2dr_+)\Omega_0^2}{r_s^2}.
\end{equation}

So, we have simplified $T_2$ to
\begin{equation}
T_2=\frac{K_4}{x-1}+\frac{K_2}{x}-\frac{\delta_r^2\Omega_0^2}{r_s^2 }.   
\end{equation}

And the simplified radial equation reads 
\begin{multline}
\partial_x^2 R+\left(\frac{1}{x-1}+\frac{1}{x}\right)\partial_{x}R+\left[\frac{\delta_r^2\left(\Omega^2-\Omega_0^2\right)}{r_s^2}\right. \\ \left.+\frac{1}{x}\left(K_2+\frac{2K_1K_3}{\delta_r}-\frac{2K_1{\Omega}}{r_s}\right)+\frac{1}{x^2}\left(\frac{{K_1}^2}{\delta_r^2}\right)+\right. \\ \left. \frac{1}{x-1}\left(K_4-\frac{2K_1K_3}{\delta_r}-\frac{2K_3\delta_r{\Omega}}{r_s}\right)+\frac{K_3^2}{{\left(x-1\right)}^2}\right]R=0
\end{multline}

Comparing with the Confluent Heun equation in \eqref{result}, we obtain exact expressions of the confluent Heun's parameters, i.e., $\alpha,\beta,\gamma,\delta$ and $\eta$ as follows,
\begin{gather}
\alpha_\pm = \pm2i\frac{\delta_r}{r_s}\sqrt{\Omega^2-\Omega_0^2}, \label{alphaeq}\\
\beta_\pm = \pm\frac{2i}{\delta_r}\bigg[\frac{\Omega}{r_s}(r_-(r_--2d)-k^2+a^2) - m_\ell a\bigg],\label{betaeq}\\
\gamma_\pm = \pm\frac{2i}{\delta_r}\bigg[\frac{\Omega}{r_s}(r_+(r_+-2d)-k^2+a^2) - m_\ell a\bigg],\label{gammaeq}\\
\delta = \frac{\delta_r}{r_s^2}(r_+ + r_- - 2d)(2\Omega^2-\Omega_0^2) \label{deltaeq},
\end{gather}
\begin{multline}
\eta =-\frac{1}{r^2_s}\left[2{\Omega}\left\{-am_\ell r_s+\left(a^2-k^2+r_-\left(r_--2d\right)\right){\Omega}\right\}\right. \\ \left.-\frac{2}{{\delta }^2_r}\left\{-am_\ell r_s+a^2{\Omega}-k^2{\Omega}+r_-\left(r_--2d\right){\Omega}\right\}\times\right. \\ \left. \left\{-am_\ell r_s+a^2{\Omega}-k^2{\Omega}+r_+\left(r_+-2d\right){\Omega}\right\}\right. \\ \left.+\left\{a^2-k^2+r_-\left(r_--2d\right)\right\}{\mathrm{\Omega }}^2_0+K^{m_\ell }_lr^2_s\right].
\end{multline}

Finally, the full exact solution of the radial Klein-Gordon equation in the region {\it inside} the inner horizon of Dyonic Kerr-Sen black hole is given by
\begin{multline}
R(r) =e^{\frac{1}{2}\alpha \left(\frac{r-r_-}{\delta_r}\right)}{\left(\frac{r-r_+}{\delta_r}\right)}^{\frac{1}{2}\gamma}\times \\
\left[A{\left(\frac{r-r_-}{\delta_r}\right)}^{\frac{1}{2}\beta}\operatorname{HeunC}\left(\alpha ,\beta ,\gamma ,\delta ,\eta ,\frac{r-r_-}{\delta_r}\right)\right. \\ \left.+B{\left(\frac{r-r_-}{\delta_r}\right)}^{-\frac{1}{2}\beta}\operatorname{HeunC}\left(\alpha ,-\beta ,\gamma ,\delta ,\eta ,\frac{r-r_-}{\delta_r}\right)\right],
\end{multline}
and the asymptotic behaviour for $|r|>>1$ is
\begin{multline}
R(r) =\frac{1}{|r|}\left[A_\infty e^{-\frac{1}{2}\alpha \left(\frac{|r|}{\delta_r}\right)}|r|^{-\frac{\delta}{\alpha}}+B_\infty e^{\frac{1}{2}\alpha \left(\frac{|r|}{\delta_r}\right)}|r|^{\frac{\delta}{\alpha}} \right].
\end{multline}

In Table I, we tabulate the signatures of $\Omega$, $\alpha$, and $\delta/\alpha$ for scalar fields inside the inner horizon of a lightly charged dyonic Kerr Sen black hole, where the condition $r_s^2>2(P^2+Q^2)$ is fulfilled.  

\begin{center}
\begin{table}[h!]
\begin{tabular}{|c|c|c|c|c|c|c|c|c|} \hline 
\multicolumn{3}{|c|}{Parameter} & $Re\left(\Omega \right)$ & $Im\left(\Omega \right)$ & $Re\left(\alpha \right)$ & $Im\left(\alpha \right)$ & $Re\left(\frac{\delta }{\alpha }\right)$ & $Im\left(\frac{\delta }{\alpha }\right)$ \\ \hline 
${\alpha }_+$ & ${\beta }_+$ & ${\gamma }_+$ & $+$ & + & - & + & - & - \\ \hline 
${\alpha }_+$ & ${\beta }_+$ & ${\gamma }_-$ & - & - & - & + & - & - \\ \hline 
${\alpha }_+$ & ${\beta }_-$ & ${\gamma }_+$ & + & + & - & + & - & - \\ \hline 
${\alpha }_+$ & ${\beta }_-$ & ${\gamma }_-$ & - & - & - & + & - & - \\ \hline 
${\alpha }_-$ & ${\beta }_+$ & ${\gamma }_+$ & - & + & + & + & + & - \\ \hline 
${\alpha }_-$ & ${\beta }_+$ & ${\gamma }_-$ & + & - & + & + & + & - \\ \hline 
${\alpha }_-$ & ${\beta }_-$ & ${\gamma }_+$ & - & + & + & + & + & - \\ \hline 
${\alpha }_-$ & ${\beta }_-$ & ${\gamma }_-$ & + & - & + & + & + & - \\ \hline 
\end{tabular}
\caption{QNMs' signatures inside inner horizon of the Dyonic Kerr Sen black hole with $r_s^2>2(P^2+Q^2)$.}
\end{table}
\end{center}

The first two rows of the table were derived by numerically calculating the quasiresonance frequencies, $\Omega$, from the exact formula we obtained for each of the eight possible modes. The resulting $\Omega$ is then used to compute the corresponding $\alpha$ of the modes, which is displayed in the third and fourth row of the table. A similar method is used to calculate $\delta/\alpha$. Notice that only $\alpha_{-}$ modes have positive Re $\left(\alpha \right)$ and the corresponding solutions are regular at $r\to -\infty$. 
    
\section{Black Hole's quasi-resonance}  \label{SecQSR}

The Confluent Heun's parameters $\alpha, \beta, \gamma$ have two possible signs in the solution. Substitute these combinations into the polynomial condition \eqref{HeunPol}, we obtain eight expressions that are functions of the Dyonic Kerr-Sen black hole parameters, i.e., mass, angular momentum, charges as well as the rest and relativistic energy of the scalar field associated with $n_r$. The polynomial condition of the radial exact solution, in fact, is simply the exact energy quantization condition, where $n_r$ is the radial quantum number. 

    \subsection{Massive Modes}
There are 8 massive modes (i.e., $2\times2\times2$ possible combinations of $\alpha_\pm,\beta_\pm$ and $\gamma_\pm$), where each solution corresponds to a quasi-resonance mode. Substituting (\ref{alphaeq})-(\ref{deltaeq}) into \eqref{HeunPol}, we obtain,

1. For mode with $\alpha_+,\beta_+,\gamma_+$, 
\begin{equation}
    \frac{(2d-r_--r_+)(2{\Omega }^2-{{{\Omega }}_0}^2)}{2r_s\sqrt{{{{\Omega }}_0}^2-{\Omega }^2}}+\frac{i}{r_s{\delta }_r}{\mathfrak{D}}_{{\Omega }}=-n_r.
\end{equation}

2. For mode with $\alpha_+,\beta_+,\gamma_-$, 
\begin{equation}
    -\frac{(2d-r_--r_+)({{{\Omega }}_0}^2-2{\Omega }^2-2\Omega \sqrt{{\Omega }^2-{{{\Omega }}_0}^2})}{2r_s\sqrt{{{{\Omega }}_0}^2-{\Omega }^2}}=-n_r.
\end{equation}

3. For mode with $\alpha_+,\beta_-,\gamma_+$, 
\begin{equation}
-\frac{(2d-r_--rp)({{{\Omega }}_0}^2-2{\Omega }^2+2\Omega \sqrt{{\Omega }^2-{{{\Omega }}_0}^2})}{2r_s\sqrt{{{{\Omega }}_0}^2-{\Omega }^2}}=-n_r.    
\end{equation}

4. For mode with $\alpha_+,\beta_-,\gamma_-$, 
\begin{equation}
\frac{(2d-r_--r_+)(2{\Omega }^2-{{{\Omega }}_0}^2)}{2r_s\sqrt{{{{\Omega }}_0}^2-{\Omega }^2}}-\frac{i}{r_s{\delta }_r}{\mathfrak{D}}_{{\Omega }}=-n_r.  
\end{equation}

5. For mode with $\alpha_-,\beta_+,\gamma_+$, 
\begin{equation}
    -\frac{(2d-r_--r_+)(2{\Omega }^2-{{{\Omega }}_0}^2)}{2r_s\sqrt{{{{\Omega }}_0}^2-{\Omega }^2}}+\frac{i}{r_s{\delta }_r}{\mathfrak{D}}_{{\Omega }}=-n_r.
\end{equation}

6. For mode with $\alpha_-,\beta_+,\gamma_-$, 
\begin{equation}
 \frac{(2d-r_--r_+)({{{\Omega }}_0}^2-2{\Omega }^2+2\Omega \sqrt{{\Omega }^2-{{{\Omega }}_0}^2})}{2r_s\sqrt{{{{\Omega }}_0}^2-{\Omega }^2}}=-n_r.   
\end{equation}

7. For mode with $\alpha_-,\beta_-,\gamma_+$, 
\begin{equation}
\frac{(2d-r_--r_+)({{{\Omega }}_0}^2-2{\Omega }^2-2\Omega \sqrt{{\Omega }^2-{{{\Omega }}_0}^2})}{2r_s\sqrt{{{{\Omega }}_0}^2-{\Omega }^2}}=-n_r.    
\end{equation}

8. For mode with $\alpha_-,\beta_-,\gamma_-$, 
\begin{equation}
-\frac{\left(2d-r_--r_+\right)\left(2{\Omega }^2-{{{\Omega }}_0}^2\right)}{2r_s\sqrt{{{{\Omega }}_0}^2-{\Omega }^2}}-\frac{i}{r_s{\delta }_r}{\mathfrak{D}}_{{\Omega }}=-n_r,
\end{equation}
where we have defined,
\begin{multline}
{\mathfrak{D}}_{{\Omega }}=-2am_\ell r_s+\left[2a^2-2k^2+r_-^2+{r_+}^2\right.\\ \left.-2d(r_-+r_+)\right]\Omega .
\end{multline}

Numerical results for certain choice of parameters for the complex scalar quasi-resonance frequencies are shown in Fig.~\ref{FIGnum}. The modes representing outgoing solutions at $r\to-\infty$ require Re $(\alpha)>0$, i.e., $\alpha_{-}$. Thus, we further eliminate all modes with $\alpha_{+}$. All of the four outgoing modes have the following properties, i.e., either growing (unstable) if having positive real frequency, or decaying if having negative real frequency. This indicates that whenever massive scalar field enters the region with CTC, the massive modes will grow over time and cause instability to the background spacetime.

 \subsection{Massless Modes}  \label{mzeroSect}
{Setting $\Omega_0 =0 $, we find another set of eight massless modes as the following,}

{
1. For mode with $\alpha_+,\beta_+,\gamma_+$, 
\begin{gather}
{{\Omega }}_1=\frac{\left[2am_\ell -i n_r\left(r_--r_+\right)\right]r_s}{2\left[a^2-k^2+r_-\left(r_--2d\right)\right]},\label{mode11}\\
{{\Omega }}_2=\frac{\left[2am_\ell -i n_r\left(r_--r_+\right)\right]r_s}{2\left[a^2-k^2+r_+\left(r_+-2d\right)\right]}.  \label{mode12}
\end{gather}}
{
2. For mode with $\alpha_+,\beta_+,\gamma_-$, 
\begin{equation}
 \Omega=\frac{i n_rr_s}{4d-2\left(r_-+r_+\right)}.  \label{mode2}
\end{equation}}
{
3. For mode with $\alpha_+,\beta_-,\gamma_+$, 
\begin{equation}
 \Omega=-\frac{i n_rr_s}{4d-2\left(r_-+r_+\right)}.   
\end{equation}}
{
4. For mode with $\alpha_+,\beta_-,\gamma_-$, 
\begin{gather}
{{\Omega }}_1=\frac{\left[2am_\ell +i n_r\left(r_--r_+\right)\right]r_s}{2\left[a^2-k^2+r_-\left(r_--2d\right)\right]},\\
{{\Omega }}_2=\frac{\left[2am_\ell +i n_r\left(r_--r_+\right)\right]r_s}{2\left[a^2-k^2+r_+\left(r_+-2d\right)\right]}    
\end{gather}}
{
5. For mode with $\alpha_-,\beta_+,\gamma_+$, 
\begin{gather}
{{\Omega }}_1=\frac{\left[2am_\ell -i n_r\left(r_--r_+\right)\right]r_s}{2\left[a^2-k^2+r_-\left(r_--2d\right)\right]},\label{mode51}\\ 
{{\Omega }}_2=\frac{\left[2am_\ell -i n_r\left(r_--r_+\right)\right]r_s}{2\left[a^2-k^2+r_+\left(r_+-2d\right)\right]}.\label{mode52}
\end{gather}}
{
6. For mode with $\alpha_-,\beta_+,\gamma_-$, 
\begin{equation}
 \Omega=\frac{i n_rr_s}{4d-2\left(r_-+r_+\right)}. \label{mode6}
\end{equation}}
{
7. For mode with $\alpha_-,\beta_-,\gamma_+$, 
\begin{equation}
 \Omega=-\frac{i n_rr_s}{4d-2\left(r_-+r_+\right)}.
\end{equation}}
{
8. For mode with $\alpha_-,\beta_-,\gamma_-$, 
\begin{gather}
{{\Omega }}_1=\frac{\left[2am_\ell +i n_r\left(r_--r_+\right)\right]r_s}{2\left[a^2-k^2+r_-\left(r_--2d\right)\right]},\label{mode81}\\
{{\Omega }}_2=\ \frac{\left[2am_\ell +i n_r\left(r_--r_+\right)\right]r_s}{2\left[a^2-k^2+r_+\left(r_+-2d\right)\right]}.\label{mode82}
\end{gather}}

\subsection{Growing Quasinormal frequency and CPC}  \label{secCPC}

In this section, we will show that all nonzero-energy massless quasinormal modes~(QNMs) within $r<r_{-}$ region are growing modes. With growing modes, the spacetime backreaction would become exponentially enhanced and cause instability of the background spacetime region where CTC exist. For the KSBH spacetime, fluctuation of massless field will destroy the spacetime region with the CTC and CPC is thus proven valid.  

First, we consider when 
\begin{equation}
a^{2}-k^{2}+r_{-}(r_{-}-2d)>0 \label{masslessgrow}.
\end{equation}
From all 8 possibilities of massless quasi-resonance modes in Sect.~\ref{mzeroSect}, only the $\alpha_{-},\beta_{+}$ modes~(see Eqn.~(\ref{betaeq})), given in Eqn.~(\ref{mode51})-(\ref{mode6}), satisfy the boundary conditions of the QNMs, namely  
\begin{eqnarray*}
&&R(r\to -\infty)\to 0, \\
&&R(r\to r_{-})\sim (\frac{r-r_{-}}{\delta_{r}})^{+i\sqrt{A_{3}}}=x^{\beta_{+}/2}.
\end{eqnarray*}  
Moreover, only the QNMs with nonzero real parts are relevant to the stability analysis of the background spacetime, the purely imaginary modes has zero energy and cannot transmit energy nor information. For all $\alpha_{-},\beta_{+}$ modes with nonzero real parts, since $r_{+}>r_{-}$ positive imaginary parts of Eqn.~(\ref{mode51})-(\ref{mode52}) naturally occur for (\ref{masslessgrow}). 

For $r_{s}^{2}>2(P^{2}+Q^{2})$, the condition (\ref{betaeq}) leads to $a^{2}>0$ after using  Eqn.~\eqref{rpm}, the BH simply has to be rotating. From real parts of the modes given in Eqn.~(\ref{mode51})-(\ref{mode52}) which is proportional to $2m_\ell a$, nonzero energy also requires the field to be circling around the BH with nonzero $m_\ell$. Thus, the nonzero energy which is the real part of QNMs and the positivity of the imaginary part of QNMs, both require the rotation of BH. 

For $r_{s}^{2}<2(P^{2}+Q^{2})$, we instead have $a^{2}-k^{2}+r_{-}(r_{-}-2d)<0$. From Eqn.~(\ref{betaeq}), in order to have the outgoing solution $r\to-\infty$ at $r_{-}$, we thus need to choose $\alpha_{-},\beta_{-}$ modes. For these modes, Im$(\Omega_{1})$  given by Eqn.~(\ref{mode81}) is apparently positive. On the other hand, Im$(\Omega_{2})$ given by Eqn.~(\ref{mode82}) will be positive if 
\begin{equation}
a^{2}-k^{2}+r_{+}(r_{+}-2d)<0 \label{masslessgrow1},
\end{equation}
which again naturally leads to the condition $a^{2}>0$, the black hole simply has to be rotating. All QNMs are also growing modes in this case. 

Growing modes lead to exponentially growing field that would generate the backreaction to the spacetime in $r<r_{-}$ region and cause instability of the spacetime. CPC is proven valid for the KSBH where region with CTC is unstable due to the exponentially growing QNMs. Numerical studies of the massive modes also found that all the outgoing positive-energy modes that can reach faraway region $x\to -\infty$ have positive imaginary parts and thus are exponentially growing modes as shown in Fig.~\ref{FIGnum}. But analysis of only massless modes is already sufficient to prove validity of CPC in the KSBH spacetime. 

A few comments on the timescale of the growing QNMs are in order. For massless modes, both Im$(\Omega_{1})$ and Im$(\Omega_{2})$ from Eqn.~(\ref{mode51})-(\ref{mode52}) and Eqn.~(\ref{mode81})-(\ref{mode82}) are proportional to $r_{+}-r_{-}$ which vanishes in the extremal limit. The QNMs approach stable normal modes in the extremal dyonic KSBH but the equation of motion becomes Double Confluent Heun equation which we will further investigate in the future work. The timescale of growing modes for {\it near-extremal} BH could thus be very long and converging together since Im$(\Omega_{1})\sim$ Im($\Omega_{2})$. 

On the other hand, Im$(\Omega_{1})$ diverges as $a\to 0$ and the timescale of the growing modes becomes arbitrarily short. Interestingly in the same $a\to 0$ limit, Im$(\Omega_{2})=n_{r}/2$ and for fundamental mode $n_{r}=1$,
\begin{equation}
{\rm Im}(\omega=\frac{c\Omega_{2}}{r_{s}})=\frac{c}{2r_{s}}=\frac{c^{3}}{4GM}.
\end{equation}
For a solar-mass BH, the timescale is $4GM_{\odot}/c^{3}=19.7$ microseconds. To illustrate the quantitative values of the timescale, for example with $r_{s}=2.0, P=0.5, Q=0.8, a=0.4$~(in geometric unit), Im $(\Omega_{1,2})\simeq 3\times 10^{8},6\times 10^{7}$ sec$^{-1}$ for fundamental modes. The characteristic time for growing and deformation is roughly nanoseconds for such KSBH.

\section{Conclusions and Discussions}
\label{SecCON}
In this work, we consider the Dyonic Kerr-Sen black hole and its spacetime structure. It is found that subextremal Dyonic Kerr-Sen black holes can be grouped into two categories; small-charge and large-charge black holes, where the latter is not possible in the Kerr-Newman BH. Comprehensive investigation of the possibilities on where the CTC may exist is then carried out and it is found that CTC always exists in the region inside the inner horizon of a black hole~(Cauchy horizon), $r<r_-$, for both categories. In addition, the contour plots of the CTCs boundaries and plots of metric functions for various combinations of the Dyonic Kerr-Sen black hole parameters are also presented. 

A novel investigation of the relativistic scalar field in the interior region of a DKSBH where CTCs exist is performed. In this work, it is for the first time the CPC is investigated by exactly solving the Klein-Gordon equation in the region inside the inner horizon of a black hole. Detail derivations to obtain the exact radial solutions of the governing Klein-Gordon equation and the quantized energy levels for both massive and massless scalar fields are presented. 

For massive fields, numerical analyses reveal that modes with $Re(\Omega)>0$ are growing over time (unstable) and modes with $Re(\Omega)<0$ are decaying, indicating whenever massive scalar fields propagate in the region with CTC, instability occurs and massive modes will grow exponentially over time as the consequence. For massless fields, instability occurs for QNMs with nonzero $Re(\Omega)$. It is analytically discovered that the condition \eqref{masslessgrow} for the instability to occur is $a^2>0$~(and $m_\ell \neq 0$), i.e., the BH simply has to be rotating and the field has to be circling around the BH. The condition, $a^2>0$, is also consistent with the condition that allows the CTC to exist (see \eqref{g33con}). The region with CTCs is thus unstable and CPC is proven valid for the KSBH. 

For other kinds of field, e.g. spinor, since we can always arrive at Klein-Gordon~(KG) equation by operating with the same Dirac operator to the Dirac equation, the solutions of such KG will be the same as (complex) scalar fields in exactly the same form as the KG equation in this work. They would contain the same growing modes which backreact and deform the spacetime region with CTC as well. The vector fields obeying generic Proca equation also have solution in the form of polarization vector multiplied to the scalar solution of KG equation and we can expect the similar exponentially growing solution in the region behind the inner horizon. 

In contrast to the widely used semi-classical WKB approximation, where the approximation breaks down for low momentum, $p=\hbar k\approx 0$, and also the breaking down of the wave function at regions with potential barriers \cite{Ashok,Boris}, or the continued fraction method, where it is efficient only for finding complex QNMs with a relatively small real part \cite{Daghigh} and does not converge for case of purely imaginary frequencies \cite{Daghigh,Moreira}. In this work, the presented exact results do not exhibit such problems and the solutions are correct for the whole region of interest. Moreover, solving the energy from the obtained quartic and quadratic equations is undoubtedly much more efficient compared to any fully numerical method since the roots are known exactly.

\begin{acknowledgments}
TB and PB are supported in part by National Research Council of Thailand~(NRCT) and Chulalongkorn University under Grant N42A660500. DS acknowledges this work is supported by the Second Century Fund (C2F), Chulalongkorn University, Thailand. 
\end{acknowledgments}

\appendix \label{AppendixA}
\section{The Confluent Heun Equation}
The Confluent Heun differential equation is a linear second order ordinary differential equation, generalization of the Hypergeometric differential equation in the canonical form \cite{Heun}, 
\begin{multline}
    \frac{d^2y_H}{dx^2}+\left(\alpha +\frac{\beta +1}{x}+\frac{\gamma +1}{x-1}\right)\frac{dy_H}{dx}
\\+\left(\frac{\mu }{x}+\frac{\nu }{x-1}\right)y_H=0, \label{canonincalheun}
\end{multline}
where,
\begin{gather}
    \mu=\frac{1}{2}\left(\alpha-\beta-\gamma+\alpha\beta-\beta\gamma\right)-\eta,\\
    \nu =\frac{1}{2}\left(\alpha +\beta +\gamma +\alpha \gamma +\beta \gamma \right)+\delta +\eta.
 \end{gather}

The solutions are given by two independent Confluent Heun functions,
\begin{multline}
    y_H=A\operatorname{HeunC}\left(\alpha ,\beta ,\gamma ,\delta ,\eta ,x\right)\\+Bx^{-\beta}\operatorname{HeunC}\left(\alpha ,-\beta ,\gamma ,\delta ,\eta ,x\right).
\end{multline}

The Confluent Heun function can be reduced to an $n^{th}$ order polynomial function if the following series termination condition is fulfilled,
\begin{equation}
\frac{\delta}{\alpha}+\frac{\beta +\gamma}{2}=-n_r,\quad n_r\in\mathbb{N}. \label{HeunPol}
\end{equation}

Suppose we have a natural general form of the asymmetrical Confluent Heun equation \cite{Heun,Vier22},
\begin{multline}
\frac{d^2y}{dx^2}+\left(\frac{1}{x}+\frac{1}{x-1}\right)\frac{dy}{dx}
\\+\left(\frac{A_1}{x}+\frac{A_2}{x-1}+\frac{A_3}{x^2}+\frac{A_4}{(x-1)^2}+A_5\right)y=0,  \label{naturalheun}
\end{multline}
in order to find the solution of \eqref{naturalheun}, we have to apply the s-homotopic transformation \cite{Alb,Heun,Vier22,Sylv,chen} by transforming the dependent variable $y(x)\to u(x)$,
\begin{equation}
y(x)=e^{B_0 x}x^{B_1}(x-1)^{B_2}u(x). \label{transf}
\end{equation}

Substituting the transformation into the equation \eqref{naturalheun}, we find the values of the exponents $B_0,B_1,B_2$ from the initial equation as follows,
\begin{gather}
B_0(B_0-1)+B_0+A_5=0 \to B_0=\pm i\sqrt{A_5},\\
B_1(B_1-1)+B_1+A_3=0 \to B_1=\pm i\sqrt{A_3},\\
B_2(B_2-1)+B_2+A_4=0 \to B_2=\pm i\sqrt{A_4}.
\end{gather}

Thus, after obtaining the three exponents, substitution of the s-homotopic transformation \eqref{transf} into \eqref{naturalheun} leads to a differential equation for $u(x)$,
\begin{multline}
    \frac{d^2u}{dx^2}+\left(2B_0+\frac{2B_1+1}{x}+\frac{2B_2+1}{x-1}\right)\frac{du}{dx}
\\+\left(\frac{\sigma}{x}+\frac{\chi}{x-1}\right)u=0,    \label{result}
\end{multline}
where
\begin{gather}
    \sigma=-B_1-B_2-2B_1B_2+B_0+2B_0B_1+A_1,\\
    \chi =B_1+B_2+2B_1B_2+B_0+2B_0B_2+A_2.
 \end{gather}

By comparing \eqref{result} with \eqref{canonincalheun}, we can write the solution for $u(x)$ in terms of the Confluent Heun functions as follows,
\begin{multline}
u=A\operatorname{HeunC}\left(\alpha ,\beta ,\gamma ,\delta ,\eta ,x\right)\\+Bx^{-\beta}\operatorname{HeunC}\left(\alpha ,-\beta ,\gamma ,\delta ,\eta ,x\right),
\end{multline}
where,
\begin{gather}
    \alpha=2B_0=\pm 2i\sqrt{A_5},\\
    \beta=2B_1=\pm 2i\sqrt{A_3},\\
    \gamma=2B_2=\pm 2i\sqrt{A_4},\\
    \delta=A_1+A_2,\\
    \eta=-A_1.
\end{gather}

Hence, the complete solutions for the natural general form of the asymmetrical Confluent Heun equation \eqref{naturalheun} are obtained as the following,
\begin{multline}
y=e^{\pm i\sqrt{A_5}x}x^{\pm i\sqrt{A_3}}(x-1)^{\pm i\sqrt{A_4}}\left[A\operatorname{HeunC}\left(\alpha ,\beta ,\gamma ,\delta ,\eta ,x\right)\right. 
\\ \left.+Bx^{-\beta}\operatorname{HeunC}\left(\alpha ,-\beta ,\gamma ,\delta ,\eta ,x\right)\right], \label{finalsol}
\end{multline}
with $\alpha,\beta,\gamma,\delta,\eta$ are given by (A.15)-(A.19).

For $x\to\infty$, the approximate solution of \eqref{canonincalheun} is given by
\begin{align}
    y_{H\infty}&=A_\infty x^{-\left(\frac{\delta}{\alpha}+\frac{\beta+\gamma+2}{2}\right)}+B_\infty e^{-\alpha x}x^{\left(\frac{\delta}{\alpha}-\frac{\beta+\gamma+2}{2}\right)}\nonumber\\
    &=e^{-\frac{1}{2}\alpha x}x^{-\frac{\beta+\gamma+2}{2}}\left[A_\infty e^{\frac{1}{2}\alpha x}x^{-\frac{\delta}{\alpha}}+B_\infty e^{-\frac{1}{2}\alpha x}x^{\frac{\delta}{\alpha}}\right],
\end{align}
or, in the form \eqref{finalsol}, we get,
\begin{equation}
    y=\frac{1}{x}\left[A_\infty e^{\frac{1}{2}\alpha x}x^{-\frac{\delta}{\alpha}}+B_\infty e^{-\frac{1}{2}\alpha x}x^{\frac{\delta}{\alpha}} \right]. \label{finalsolinfty}
\end{equation}

\begin{figure*}[h!]
\centering
\includegraphics[width=16.5cm]{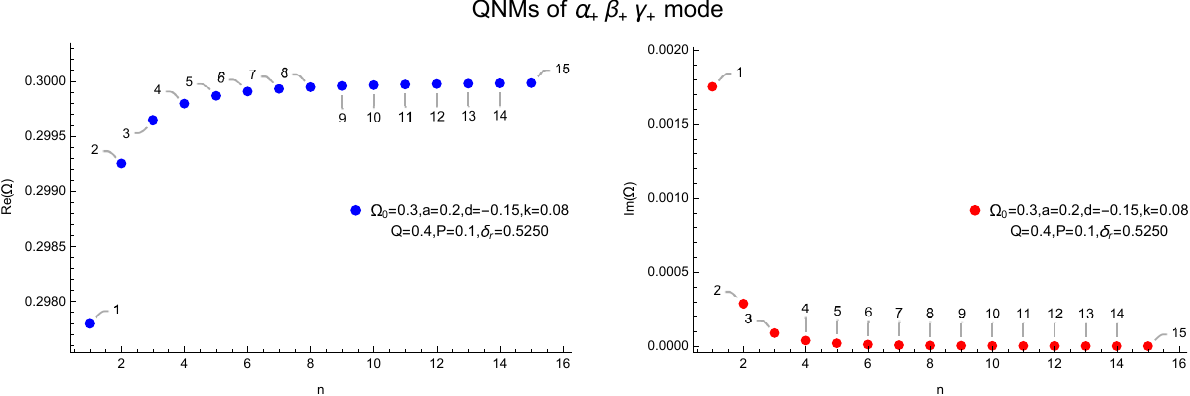}
\end{figure*}
\begin{figure*}
\centering
\includegraphics[width=16.5cm]{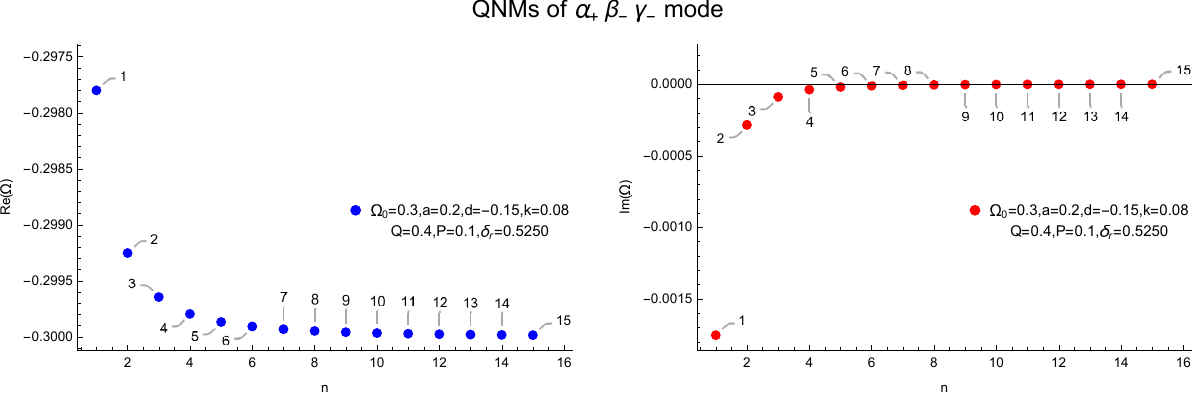}
\end{figure*}
\begin{figure*}
\centering
\includegraphics[width=16.5cm]{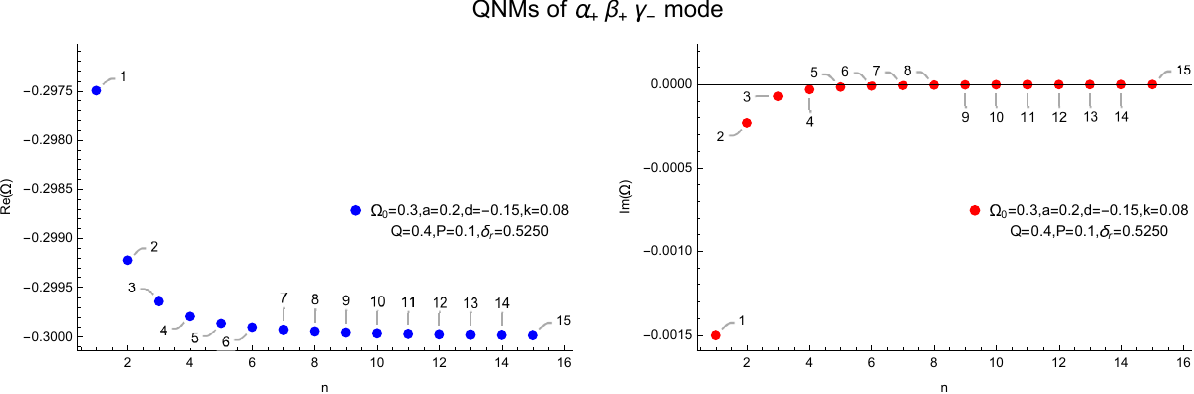}
\end{figure*}
\begin{figure*}
\centering
\includegraphics[width=16.5cm]{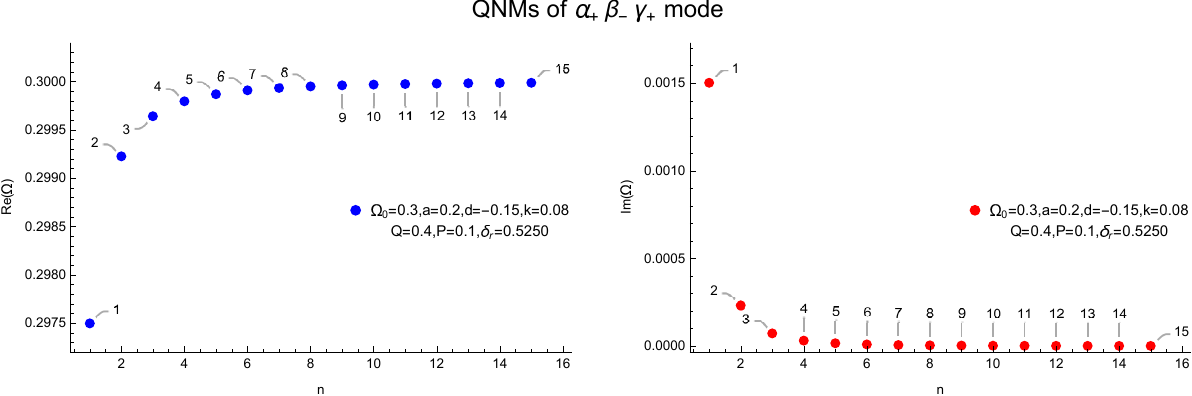}
\end{figure*}
\begin{figure*}
\centering
\includegraphics[width=16.5cm]{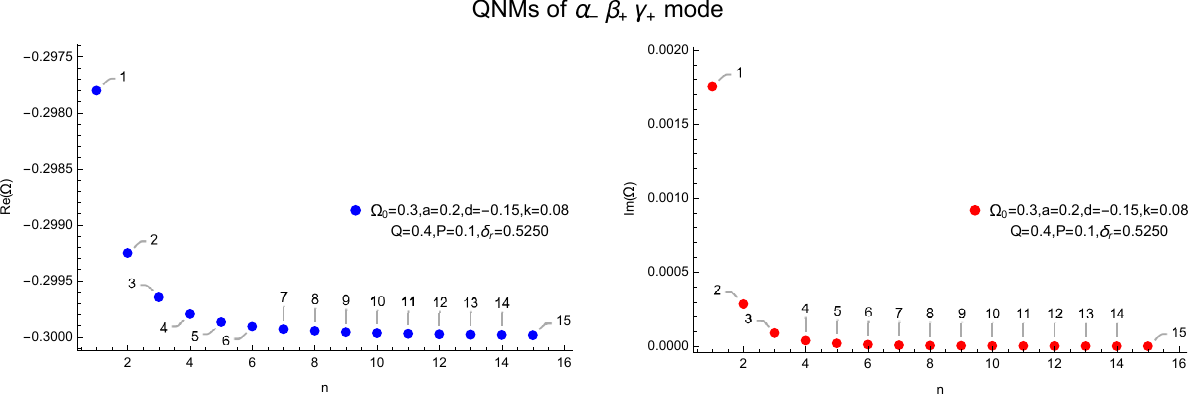}
\end{figure*}
\begin{figure*}
\centering
\includegraphics[width=16.5cm]{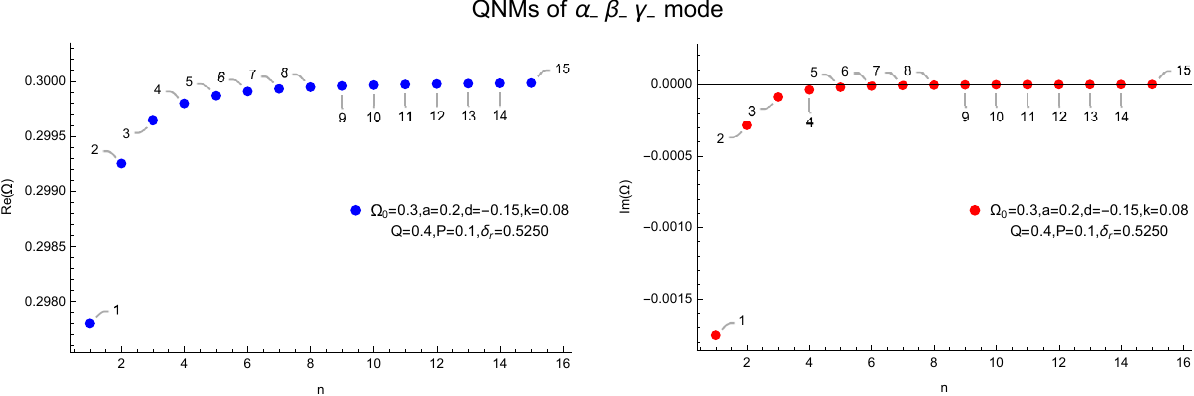}
\end{figure*}
\begin{figure*}
\centering
\includegraphics[width=16.5cm]{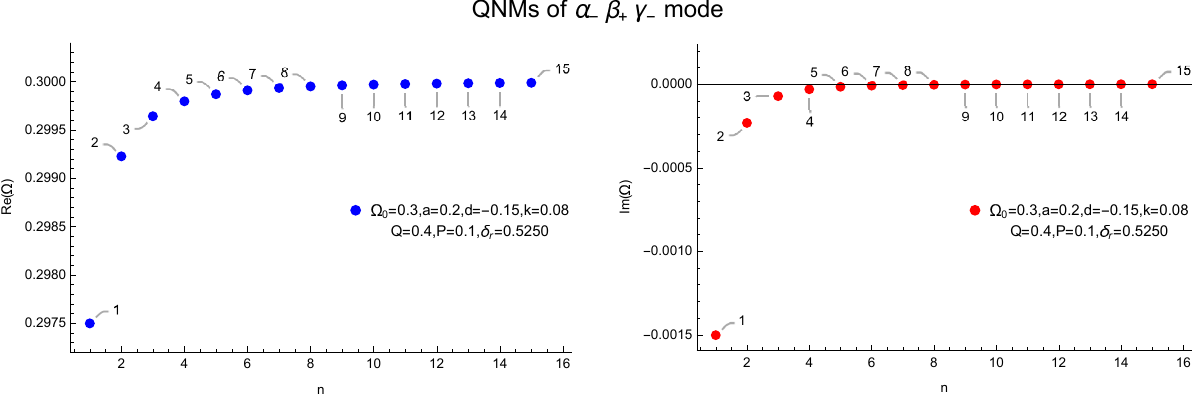}
\end{figure*}
\begin{figure*}
\centering
\includegraphics[width=16.5cm]{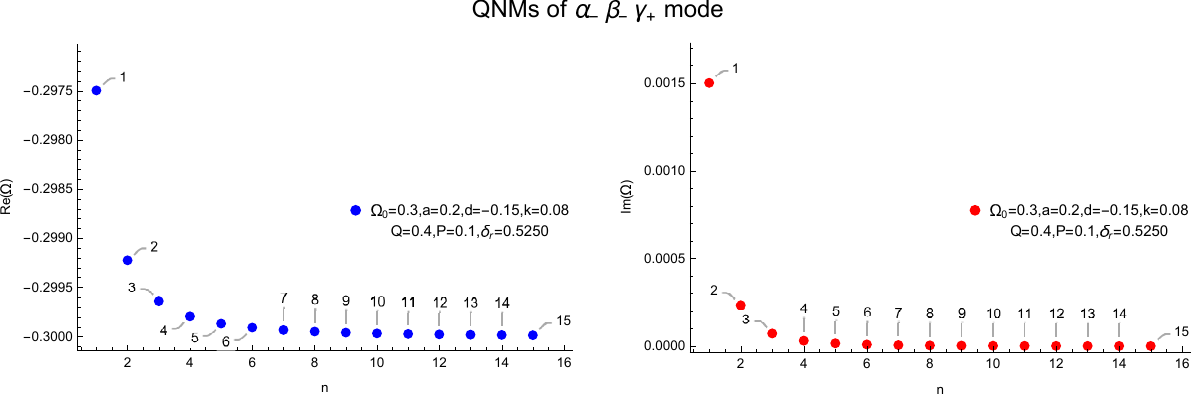}
\caption{Massive QNM's frequencies for all modes with $\Omega_{0}=0.3$.}\label{FIGnum}
\end{figure*} 

\end{document}